\documentclass[%
superscriptaddress,
reprint,
 amsmath,amssymb,
 aps,
 pre,
]{revtex4-2}

\usepackage{epsfig,gensymb,units,textcomp,color,leftidx,tabularx}
\usepackage[acronym]{glossaries}

\usepackage{lipsum}

\usepackage{graphicx}
\usepackage{dcolumn}
\usepackage{bm}
\usepackage[caption=false]{subfig}
\usepackage{xspace}
\usepackage{hyperref}
\newacronym{LMJ}{LMJ}{Laser MegaJoule}
\newcommand{\lmj}{\gls{LMJ}\xspace}
\newacronym{NIF}{NIF}{National Ignition Facility}
\newcommand{\NIF}{\gls{NIF}\xspace}
\newacronym{TCC}{TCC}{Target Chamber Center}
\newcommand{\TCC}{\gls{TCC}\xspace}
\newacronym{TNSA}{TNSA}{Target Normal Sheath Acceleration}
\newcommand{\TNSA}{\gls{TNSA}\xspace}
\newacronym{CBET}{CBET}{Cross-Beam Energy Transfer}
\newcommand{\CBET}{\gls{CBET}\xspace}
\newacronym{HED}{HED}{High Energy Density}
\newcommand{\HED}{\gls{HED}\xspace}
\newacronym{MHD}{MHD}{magnetohydrodynamics}
\newcommand{\MHD}{\gls{MHD}\xspace}
\newacronym{ICF}{ICF}{Inertial Confinement Fusion}
\newcommand{\ICF}{\gls{ICF}\xspace}
\newacronym{MagLIF}{MagLIF}{Magnetized Liner Inertial Fusion}
\newcommand{\maglif}{\gls{MagLIF}\xspace}
\newacronym{MIFEDS}{MIFEDS}{magneto-inertial fusion electrical discharge system}
\newcommand{\mifeds}{\gls{MIFEDS}\xspace}
\newacronym{XRFC}{XRFC}{X-Ray Framing Cameras}
\newcommand{\xrfc}{\gls{XRFC}\xspace}
\newacronym{RCF}{RCF}{radiochromic film}
\newcommand{\rcf}{\gls{RCF}\xspace}
\newacronym{SRS}{SRS}{Stimulated Raman Scattering}
\newcommand{\SRS}{\gls{SRS}\xspace}
\newacronym{TPD}{TPD}{Two Plasmon Decay}
\newcommand{\TPD}{\gls{TPD}\xspace}
\newacronym{LPI}{LPI}{Laser-Plasma Instabilities}
\newcommand{\lpi}{\gls{LPI}\xspace}
\newacronym{LDCs}{LDCs}{Laser-Driven Coil targets}
\newcommand{\LDC}{\gls{LDCs}\xspace}

\newcommand{\LLNL}{Lawrence Livermore National Laboratory, Livermore, California 94550, USA}
\newcommand{\CELIA}{Université de Bordeaux-CNRS-CEA, Centre Lasers Intenses et Applications (CELIA), UMR 5107, F-33405 Talence, France}
\newcommand{\CEA}{CEA, DAM, DIF, F-91297 Arpajon, France}
\newcommand{\CESTA}{CEA-CESTA, CS 60001, 33116 Le Barp Cedex, France}
\newcommand{\LULI}{LULI-CNRS, CEA, Sorbonne Universites, Ecole Polytechnique, Institut Polytechnique de Paris, F-91128 Palaiseau Cedex, France}
\newcommand{\UCSD}{Center for Energy Research, University of California-San Diego, La Jolla, CA 92093, United States of America}
\newcommand{\UAlberta}{Department of Electrical and Computer Engineering, University of Alberta, Edmonton, T6G1R1 Alberta, Canada}

\newcommand{\York}{Department of Physics, University of York, Heslington YO10 5DD, United Kingdom}
\newcommand{\CLPU}{Centro de Laseres Pulsados, Building M5, Science Park, 37185 Villamayor, Salamanca,
Spain}

\newcommand{\UPM}{ETSI Aeron\'autica y del Espacio, Universidad Polit\'ecnica de Madrid, 28040 Madrid, Spain}
\newcommand{\PIIM}{Aix Marseille Université, CNRS, PIIM, F-13013 Marseille, France}
\newcommand{\ULPGC}{iUNAT–Departamento de F\'isica, Universidad de Las Palmas de Gran Canaria, 35017 Las Palmas de Gran Canaria, Spain}
\newcommand{\UVa}{Departamento de Física Te\'orica, At\'omica y  \'Optica, Universidad de Valladolid, 47011 Valladolid, Spain}
\newcommand{\UReno}{Department of Physics, University of Nevada, Reno, Nevada 89557, USA}

\newcommand{\GA}{General Atomics, San Diego, California 92121, USA.}

\newcommand{\IPPL}{Institute of Plasma Physics $\&$ Lasers, Hellenic Mediterranean University Research Centre, 74100 Rethymno, Greece}
\newcommand{\ICL}{Plasma Physics Group, The Blackett Laboratory, Imperial College London, London, SW7 2AZ, UK}
\newcommand{\ELI}{ELI-Beamlines, Institute of Physics, Czech Academy of Sciences, 25241 Doln\'i Brezany, Czech Republic}

\begin{document}

\title{A cylindrical implosion platform for the study of highly magnetized plasmas at LMJ}

\author{G. P\'erez-Callejo}
	\thanks{gabriel.perez.callejo@uva.es}
	\affiliation{\CELIA}
	\affiliation{\UVa}
\author{C. Vlachos}
	\affiliation{\CELIA}
	\affiliation{\IPPL}
\author{C. A. Walsh}
	\affiliation{\LLNL}
\author{R. Florido}
    \affiliation{\ULPGC}
\author{M. Bailly-Grandvaux}
    \affiliation{\UCSD}
\author{X. Vaisseau}
    \affiliation{\CEA}
\author{F. Suzuki-Vidal}
    \affiliation{\ICL}
\author{C. McGuffey}
    \affiliation{\GA}
\author{F. N. Beg}
    \affiliation{\UCSD}
\author{P. Bradford}
    \affiliation{\CELIA}
\author{V. Ospina-Boh\'orquez}
    \affiliation{\CELIA}
    \affiliation{\CEA}
    \affiliation{University of Salamanca, 37008 Salamanca, Spain}
    \affiliation{Universit\'e Paris-Saclay, CEA, LMCE, 91680 Bruy\`eres-le-Ch\^atel, France}
\author{D. Batani}
    \affiliation{\CELIA}
\author{D. Raffestin}
    \affiliation{\CELIA}
\author{A. Cola\"itis}
    \affiliation{\CELIA}
\author{V. Tikhonchuk}
    \affiliation{\CELIA}
    \affiliation{\ELI}
\author{A. Casner}
    \affiliation{\CELIA}
    \affiliation{\CESTA}
\author{M. Koenig}
    \affiliation{\LULI}
\author{B. Albertazzi}
    \affiliation{\LULI}
\author{R. Fedosejevs}
    \affiliation{\UAlberta}
\author{N. Woolsey}
    \affiliation{\York}
\author{M. Ehret}
    \affiliation{\CLPU}
\author{A. Debayle}
    \affiliation{\CEA}
    \affiliation{Universit\'e Paris-Saclay, CEA, LMCE, 91680 Bruy\`eres-le-Ch\^atel, France}
\author{P. Loiseau}
    \affiliation{\CEA}
    \affiliation{Universit\'e Paris-Saclay, CEA, LMCE, 91680 Bruy\`eres-le-Ch\^atel, France}
\author{A. Calisti}
    \affiliation{\PIIM}
\author{S. Ferri}
    \affiliation{\PIIM}
\author{J. Honrubia}
    \affiliation{\UPM}
\author{R. Kingham}
    \affiliation{\ICL}
\author{R. C. Mancini}
    \affiliation{\UReno}
\author{M. A. Gigosos}
    \affiliation{\UVa}
\author{J. J. Santos}
	\thanks{joao.santos@u-bordeaux.fr}
	\affiliation{\CELIA}

\date{\today}


\begin{abstract}
Investigating the potential benefits of the use of magnetic fields in Inertial Confinement Fusion (ICF) experiments has given rise to new experimental platforms like the Magnetized Liner Inertial Fusion (MagLIF) approach at the Z-machine (Sandia National Laboratories), or its laser-driven equivalent at OMEGA (Laboratory for Laser Energetics). Implementing these platforms at MJ-scale laser facilities, such as the Laser MegaJoule (LMJ) or the National Ignition Facility (NIF), is crucial to reaching self-sustained nuclear fusion and enlarges the level of magnetization that can be achieved through a higher compression. In this paper, we present a complete design of an experimental platform for magnetized implosions using cylindrical targets at LMJ. A seed magnetic field is generated along the axis of the cylinder using laser-driven coil targets, minimizing debris and increasing diagnostic access compared with pulsed power field generators. We present a comprehensive simulation study of the initial B-field generated with these coil targets, as well as 2-dimensional extended magneto-hydrodynamics (MHD) simulations showing that a \unit[5]{T} initial B-field is compressed up to \unit[25]{kT} during the implosion. Under these circumstances, the electrons become magnetized, which severely modifies the plasma conditions at stagnation. In particular, in the hot spot the electron temperature is increased (from \unit[1]{keV} to \unit[5]{keV}) while the density is reduced (from $\unit[40]{g/cm^{3}}$ to $\unit[7]{g/cm^{3}}$). We discuss how these changes can be diagnosed using X-ray imaging and spectroscopy, and particle diagnostics. We propose the simultaneous use of two dopants in the fuel (Ar and Kr) to act as spectroscopic tracers. We show that this introduces an \textit{effective spatial resolution} in the plasma which permits an unambiguous observation of the B-field effects. Additionally, we present a plan for future experiments of this kind at LMJ.
\end{abstract}

\maketitle

\section{Introduction}
\label{sec:introduction}

Magnetization is a promising strategy to increase fusion yields and relax ignition criteria in laser-driven \ICF \cite{perkins2013}, as the presence of a B-field strongly modifies fundamental properties of \HED plasmas. In \ICF implosions, this concerns (among other key mechanisms) heat transport, which governs the transfer of the laser energy from the corona to the ablation front, and becomes anisotropic \cite{hill2018,hill2021} in the presence of a strong B-field ($\sim$\unit[kT]). Seed B-fields can be amplified by $\sim 500$ times to strengths up to $B>\unit[10]{kT}$ through compression of the plasma \cite{chang2011}. This increases the fusion yields as it inhibits the thermal energy transport and reduces the loss of $\alpha$ particles from the hotspot perpendicularly to the B-field \cite{jones1986,basko2000,chang2011,perkins2013}. Additionally, magnetized implosions may be less vulnerable to hydrodynamic instabilities \cite{sano2013,walsh2022} that could lead to disadvantageous mixing of the hot and cold parts of the target \cite{walsh2019}. Suppressing these instabilities is crucial issue to reach ignition at the \NIF \cite{kline2019, kritcher2022, zylstra2022}.

A common approach to magneto-inertial fusion is the use of cylindrical geometries with an axial B-field, which was originally used in the Z-pinch \maglif experiments at the Z-machine \cite{slutz2010, gomez2014}. Profiting from the advantages described above, combined with this favorable geometry, the cylindrical compression is expected to be near-adiabatic and stable, with a lower implosion velocity and convergence ratio than in conventional \ICF. To explore this approach with a higher repetition rate and easier diagnostic access than on the Z-machine, a laser-driven downscaled \maglif approach is being explored at the OMEGA 60 laser facility \cite{OMEGA_Paper}, thus facilitating investigations of the underlying physics \cite{davies2017, barnak2017, hansen2018a, hansen2018b, hansen2020}. At OMEGA, the axial seed B-field is generated using external capacitive pulsed discharges with the \mifeds \cite{gotchev2009}, which can produce B-fields of up to $\sim \unit[30]{T}$.

However, the \mifeds system blocks the line-of-sight that follows the axis of the cylinder, which complicates the study of radial gradients and instabilities during the implosion. Additionally, it produces a significant amount of debris that can be damaging to the nearby diagnostics and the facility in general. As an alternative strategy, the generation of B-fields using \LDC has recently begun to be investigated for cylindrical implosion experiments at OMEGA \cite{mcguffey2020,bailly2020}. This approach builds upon the demonstration of the compactness and performance of these targets observed in recent laser-plasma experiments carried out with in laser facilities of more modest energies \cite{santos2015, law2016, tikhonchuk2017, santos2018, bailly2018}. Laser-driven B-field targets could potentially extend the range of available magnetization levels that can be reached in the imploding plasma. Additionally, contrary to \mifeds, they are practically debris-free, and do not considerably block the line of sight of diagnostics to the target, thus facilitating the study of the implosion in both axial and radial directions. Independently of the method that is used to generate the seed B-field, the extreme magnetization phenomena produced in these cylindrical implosion experiments have been recently studied using extended \MHD simulations \cite{walsh2021}.

In this context we present an experimental design for studying the dynamics of imploding plasmas under extreme magnetizations at the \lmj facility \cite{casner2015lmj, miquel2016laser}. The platform described here is scheduled to be fielded in 2024-2026. We build on the aforementioned \maglif experiments at OMEGA 60 using \LDC. This paper constitutes a first report on these experiments. 

We present results from using state-of-the-art \MHD, atomic physics and radiation transport simulation tools, which show that it is possible to reach conditions for extreme magnetizations using a relatively low seed B-field. In particular, a \unit[5]{T} initial B-field can be compressed to $\unit[>10]{kT}$, thus allowing unique studies of the impact of magnetization on electron heat transport, magnetic flux compression, stagnation temperatures and fusion reaction yields. 

The paper is structured as follows. In section \ref{sec:ExpConf} the experimental configuration is described, including the restrictions for target dimensions, and laser pointing. Section \ref{sec:coilModel} presents the expected values of the seed B-field that can be produced within this design. This is expanded upon in the Appendix, which describes the physical model used to describe the B-field generation using \LDC and details the results and analysis from a recent experiment at OMEGA in which \LDC were characterized in conditions similar to those at \lmj. In section \ref{sec:Results}, we present the results from extended \MHD simulations showing the compression of the seed B-field and its effect on the plasma conditions compared to the simulations for the unmagnetized case. Section \ref{sec:diagnostic} discusses a variety of diagnostics to characterize the experiment, and finally, section \ref{sec:Conclusions} discusses the conclusions and future perspectives of this work.

\section{Experimental configuration}
\label{sec:ExpConf}

\begin{figure}
    \centering
    \includegraphics[width=0.8\columnwidth]{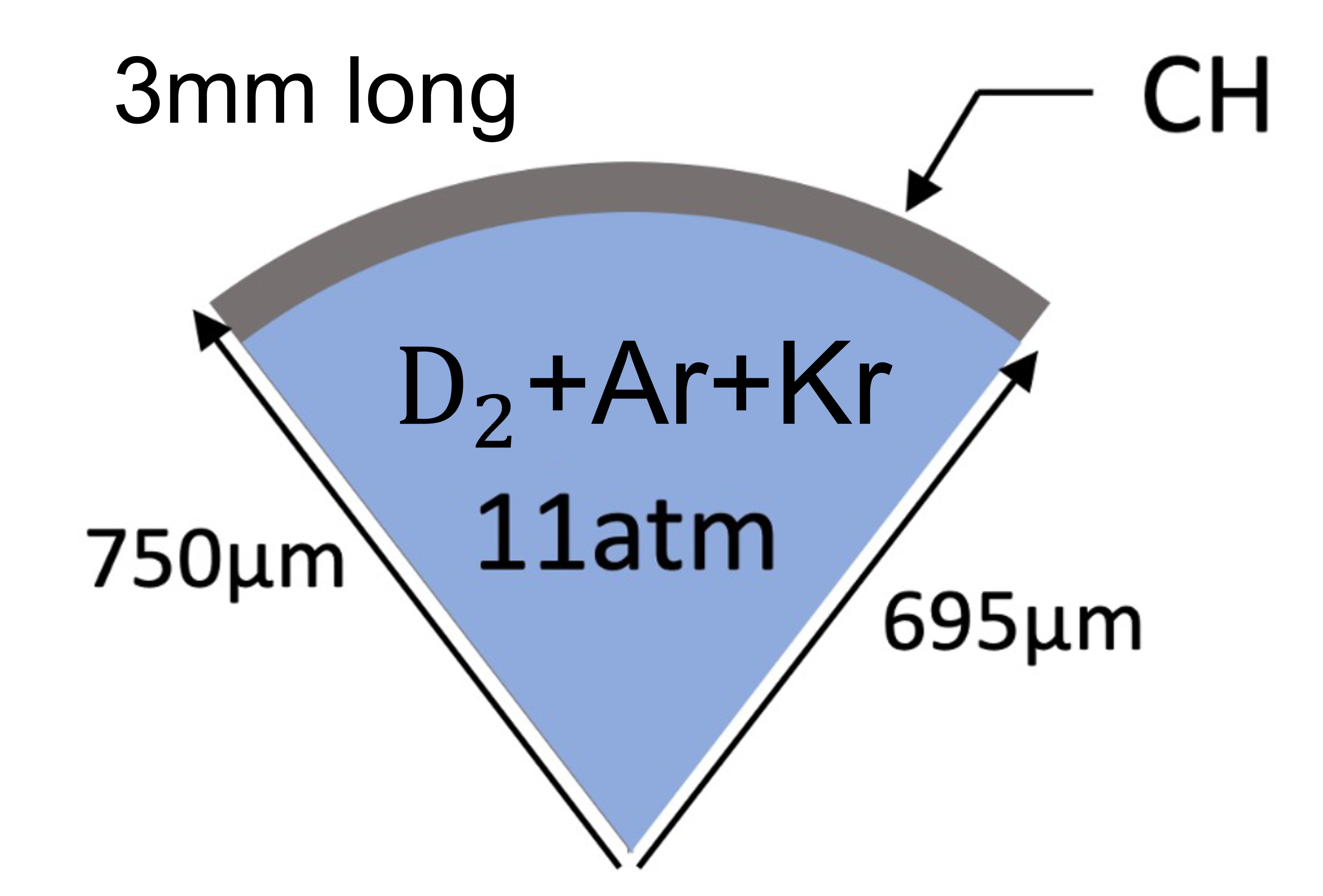}
    \caption{Schematic design of the cylindrical targets, showing the plastic shell and the gas fuel. This cylinder is \unit[3]{mm} long, and the thickness of the CH shell is $\unit[55]{\micro m}$.}
    \label{fig:target_design}
\end{figure}

\begin{figure*}
\subfloat[\label{fig:visrad_full}]{
    \includegraphics[height=0.35\textwidth]{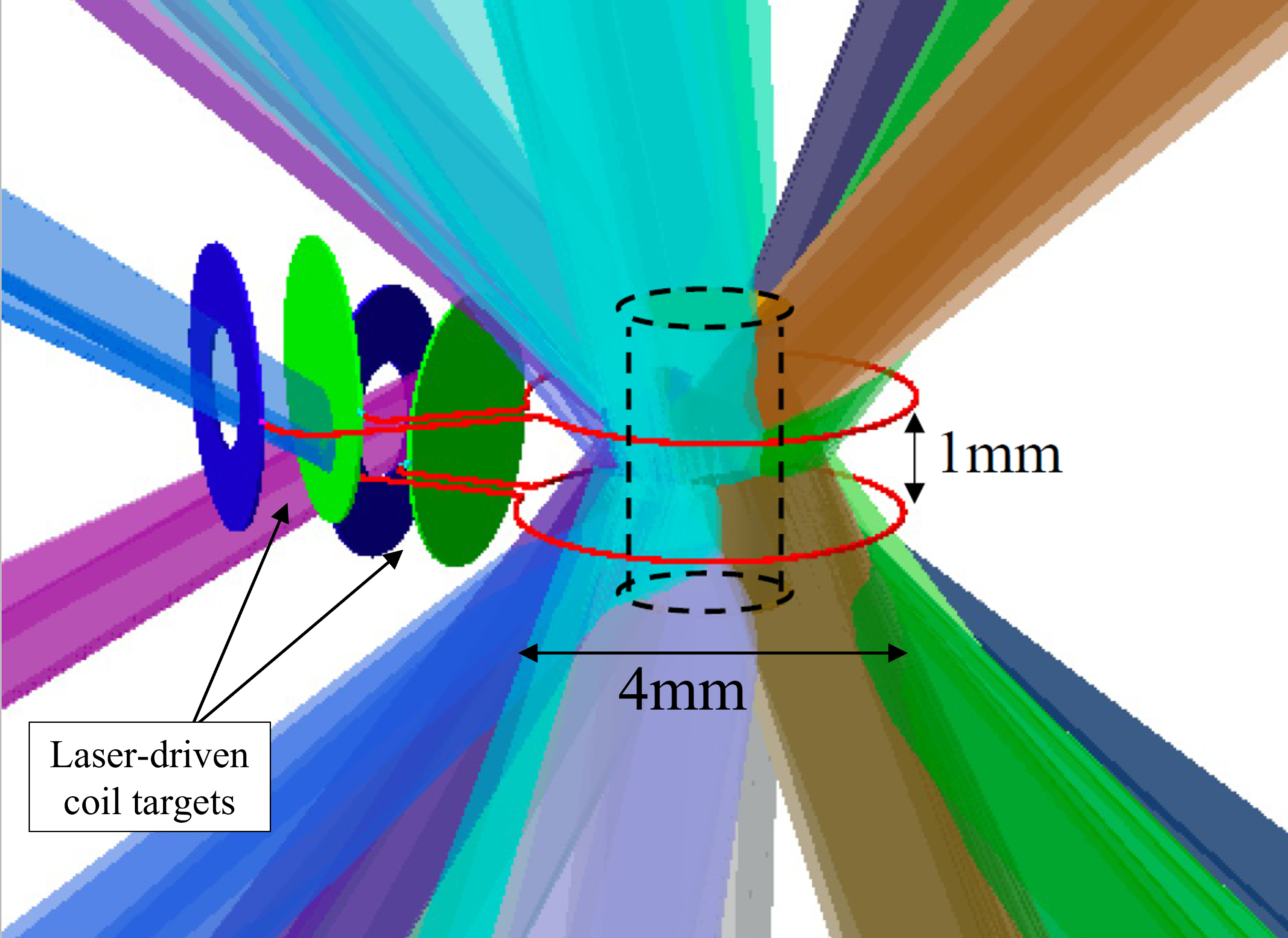}
    }
  \hfill
\subfloat[\label{fig:laser_imprint}]{
	\includegraphics[height=0.35\textwidth]{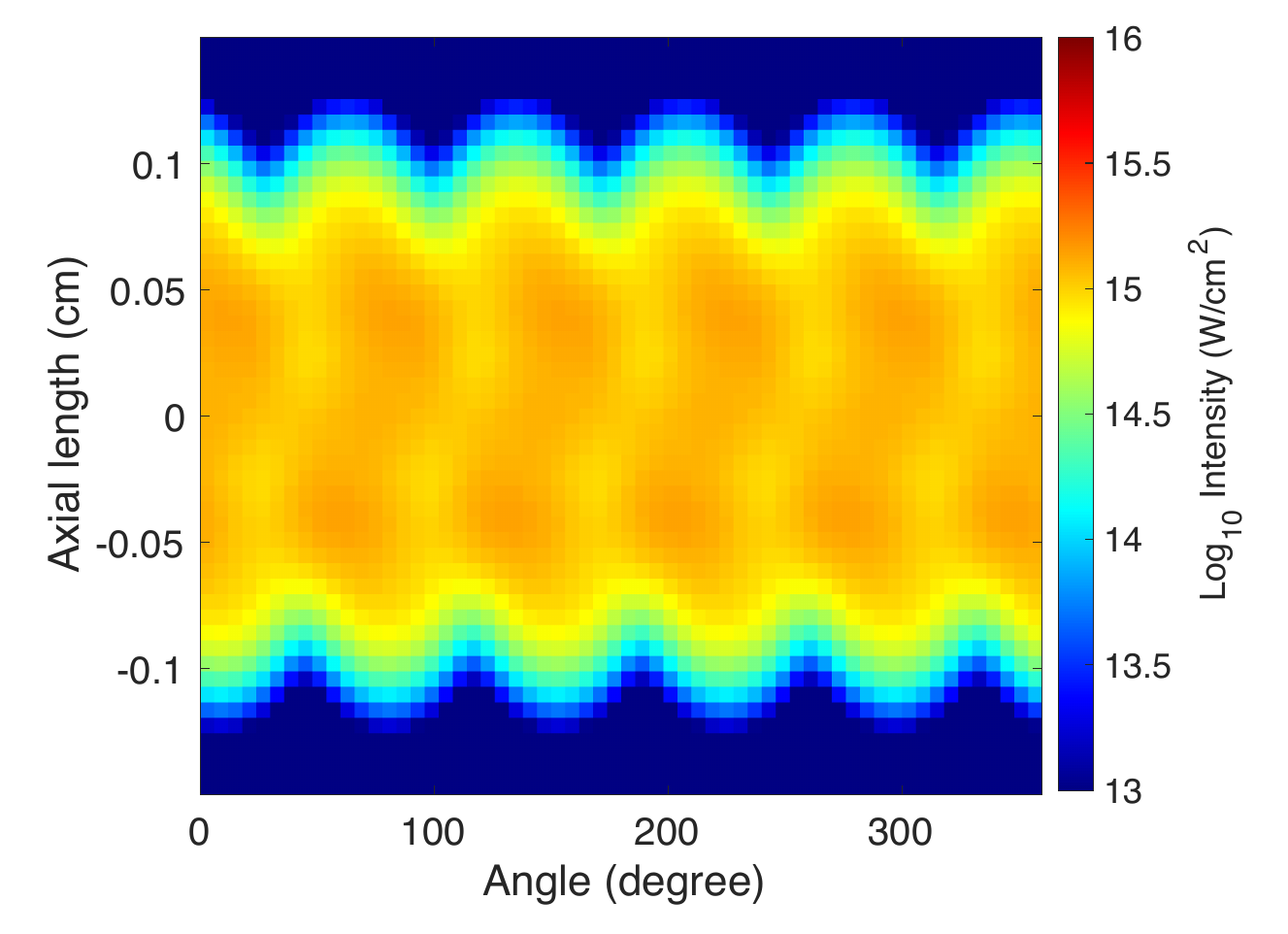}
}
\caption{(\ref{fig:visrad_full}) VisRad \cite{visrad} image showing the proposed configuration of the targets, with the 80 active laser beams for cylindrical compression. The beams' radius in display corresponds to 99\% enclosed beam energy. The cylindrical target position is displayed by a black dashed line contour. Additionally, the figure shows the coil targets, mounted around the cylinder, with a gap distance between coils of \unit[1]{mm}. In this configuration, the B-field generated at the axis of the cylinder is expected to be $\sim$\unit[5-10]{T}. It can be seen how the coil targets are away from the \lmj beams that irradiate the cylinder. (\ref{fig:laser_imprint}) Laser intensity on the cylindrical target using the proposed drive configuration. The intensity is $\unit[(11.6\pm1.2)\times 10^{14}]{W cm^{-2}}$ over the central \unit[1]{mm} region.}
\end{figure*}

The proposed experimental configuration consists of a D$_2$-filled plastic cylinder positioned at the \TCC, whose axis is aligned with the target chamber vertical axis. The cylinder is \unit[3]{mm} long, $\unit[750]{\micro m}$ outer radius – $\unit[55]{\micro m}$ thick plastic shells (CH, $\unit[1.1]{g/cm^{3}}$) filled with D$_2$ at \unit[11]{atm} ($\unit[1.81]{mg/cm^{3}}$), as shown in Figure \ref{fig:target_design}. Targets have been designed to have the minimum size, while ensuring that the focused \lmj laser beams are effectively terminated on target.

We propose the use of a dopant in the gas to regulate the core temperature \cite{walsh2021} and, most importantly, to act as a spectroscopic tracer for characterizing the plasma conditions. In particular, we propose using argon, krypton, or rather a combination of the two. Both elements have been previously used as spectroscopic tracers for \ICF-related experiments \cite{florido2008, florido2011, florido2014, chen2017, gao2022}. While the dopant concentration will decrease the obtainable temperature owing to radiative losses, the choice of dopant will determine the plasma conditions that can be probed using spectroscopic diagnostics. This is discussed further in Section \ref{sec:diagnostic}.

The cylindrical target is imploded using 80 laser beams as shown in Figure \ref{fig:visrad_full}. These are grouped into twenty groups (quads), each of which delivers a total energy of \unit[13.5]{kJ} using a \unit[3]{ns} square pulse \cite{CourtoisLMJ} of $3\omega$ light ($\lambda=\unit[351]{nm}$). The quads are uniformly distributed in four rings around the vertical axis of the chamber, with polar angles 33.2\degree, 49\degree, 131\degree and 146.8\degree. This generates an irradiation profile on the cylinder as shown in Figure \ref{fig:laser_imprint}. Over the central region of the target ($\pm \unit[0.5]{mm}$), the drive is reasonably uniform, with an intensity of $\unit[(11.6\pm 1.2)\times 10^{14}]{Wcm^{-2}}$. This corresponds to a 6\% variation in the azimuthal direction, and a 4\% axial variation.

Around the cylindrical target, two copper coil targets in a quasi-Helmholtz configuration are mounted, as shown in Figure \ref{fig:visrad_full}. The purpose of these laser-driven coils is to generate a seed B-field along the axis of the cylinder \cite{zhu2015, fiksel2016, goyon2017, tikhonchuk2017, santos2018, peebles2020, morita2021}, and magnetize the central region of the fuel. The coils are positioned at $z=-\unit[0.5]{mm}$ and $z=\unit[0.5]{mm}$, where $z=0$ corresponds to \TCC (thus covering the central \unit[1]{mm} axial length of the cylinder). The coils' axes are coincident with the cylinder target axis and with the vertical axis of the interaction chamber. Although the exact design of the coil targets (diameter of the plates, wire length and orientation) can be modified, the diameter of the coils must be large enough so that they are not irradiated by the laser beams driving the implosion (in Figure \ref{fig:visrad_full}, we show the radius of the laser beams that contains 99\% of their energy). We choose a coil diameter of \unit[4]{mm}, to minimize the target inductance,yielding \unit[13]{nH}, while keeping clear of the beams. 

Recent experiments at the PALS \cite{PALSpaper} and LULI \cite{LULIpaper} laser facilities using laser-driven coil targets indicate that the shock generated in the irradiated plate takes $\sim\unit[1]{ns}$ to traverse the thickness of a \unit[50]{\micro m}-thick plate (in agreement with hydrodynamic simulations), and that the X-ray radiation emitted from the back of the plate is too weak to pre-heat targets on the other side. Nevertheless, as an additional precaution, in order to avoid direct X-ray irradiation from these copper plates that could modify the implosion dynamics, in the design presented here the plates of the coil targets are oriented so that their surface normal directions do not intersect with the cylindrical target. The side-on emission from the plasma generated between the plates does not intersect with the cylindrical target either. Besides, the thickness of the irradiated plate must be adapted to the laser pulse duration.

The coil targets are driven by two additional $3\omega$ laser quads (one per coil target), delivering a total of \unit[13.5]{kJ} to each coil target, with a circular focal spot, $\unit[375]{\micro m}$ in diameter (at $1/e$ intensity), and a minimum pulse length of \unit[3]{ns}. This corresponds to a maximum intensity of $\sim\unit[4\times 10^{15}]{Wcm^{-2}}$ on the coil plates, although the duration of the pulse can be extended to tune the initial B-field. These quads are incident at 59.5\degree ~and 120.5\degree ~from the vertical axis of the chamber respectively and separated by an azimuthal angle of 18\degree.

\section{Generation of the seed B-field}
\label{sec:coilModel}

Each of the \LDC consists of two plates, marked in green and blue in Figure \ref{fig:visrad_full}, connected by a single-loop coil. Laser beams are focused onto the inner plate (green in the figure), thereby ejecting hot electrons towards the outer plate (blue in the figure). This process determines the efficiency of generating a current in the coil and the induced B-field.

The underlying physics of \LDC can be described by the plasma \textit{diode} model developed by Tikhonchuk \textit{et al.} \cite{tikhonchuk2017}, where the coil-targets are considered as a resistor-inductor (RL) circuit fed by a laser-driven diode current. The laser is typically of ns-duration and sufficiently intense to generate a significant number of non-thermal (hot) electrons. The hot electron temperature is obtained as a function of the laser irradiance $I\lambda^2$, using known scaling laws \cite{tikhonchuk2017}. With respect to the conversion efficiency of laser energy into hot electrons, values in the literature range from $\sim 1\%$ \cite{hohenberger2014, renner2016, cristoforetti2017, batani2018, zhang2020} to $\sim 10\%$ \cite{beg1997, price1995}. Here we assume a conversion efficiency of $1\%$, following the recent results from Zhang \textit{et al.} \cite{zhang2020}, obtained at the OMEGA facility for $3\omega$ laser interactions with solid foils. Further details on this model and the physics behind the \LDC, are given in the Appendix.

\begin{figure}
    \centering
    \includegraphics[width=\columnwidth]{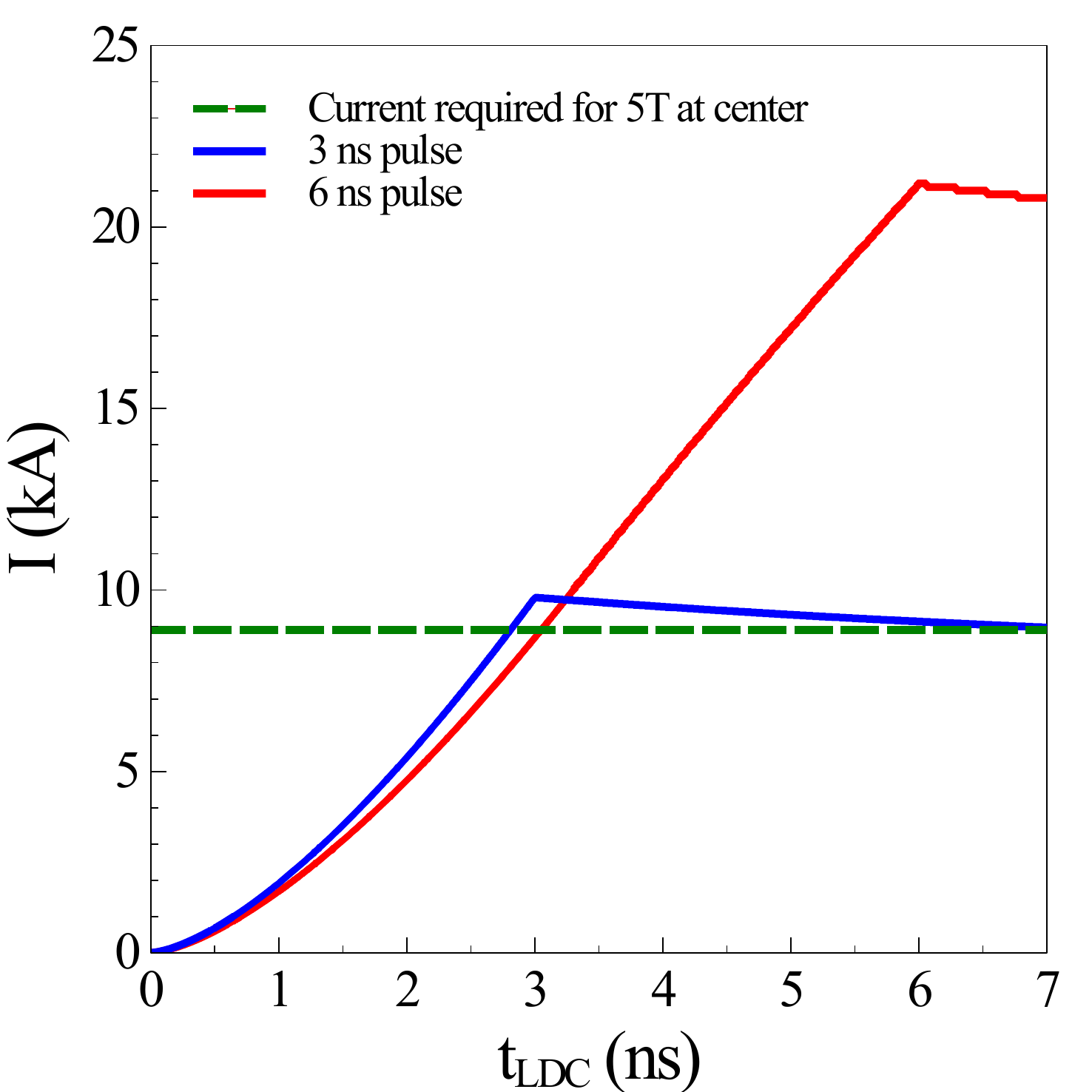}
    \caption{Time evolution of the current looping through the coil targets for this experiment, depending on the duration of the laser pulse. The timescale corresponds to that of the laser beams driving the \LDC. The main target beams are delayed by 3 and \unit[6]{ns} respectively, so that the seed B-field is maximum prior to their activation.}
    \label{fig:Current_LMJ}
\end{figure}

For the conditions at \lmj the predicted hot electron temperatures are \unit[7.5]{keV} and \unit[6.5]{keV}, for 3 and \unit[6]{ns} laser pulse drives, respectively. Correspondingly, we estimate peak currents on the coils of \unit[10]{kA} and \unit[20]{kA}. These correspond to a seed B-field between 5.5 and $\sim$\unit[10]{T} at the center of the quasi-Helmholtz system. For the rest of this paper we will work with a conservative value of \unit[5]{T}, to ensure the feasibility of the platform. 

The inductance of the proposed \LDC for \lmj is \unit[13]{nH}, twice the value of those used at OMEGA (see Appendix). This is a limiting constraint determined by the large spatial scale of the experiment. However, the configuration of the \lmj beams allows to place the two coils in Helmholtz configuration, as shown in Figure \ref{fig:visrad_full}, which allows to amplify the B-field over the cylinder volume.

Figure \ref{fig:Current_LMJ} shows the time evolution of the current for the two laser pulse durations mentioned above (3 and \unit[6]{ns}), together with the current that is required for a \unit[5]{T} field at the center of the Helmholtz system (\unit[8.9]{kA}). Note that, although the peak current is obtained at the end of the laser pulse, owing to the RL-circuit behavior, after the end of the laser the current decreases very slowly, staying relatively constant over several nanoseconds. The beams that drive the cylindrical target are activated at the end of the LDC drive, in order to reach the maximum seed B-field before the main target is imploded.

Although the highest B-field is obtained for a longer pulse since the characteristic RL time is still longer than the laser drive, it should be noted that the pulse cannot be arbitrarily long, since there are additional physical processes that will limit the current. If the pulse is too long, the generated shock will break out on the opposite side of the driven plate, thus leading to a reduction of the laser-target energy coupling. Additionally, if the plasma between the plates becomes too dense, it may short-circuit the coil. Besides, the laser energy will no longer be deposited in the rear plate.

\begin{figure}
\includegraphics[width=\columnwidth]{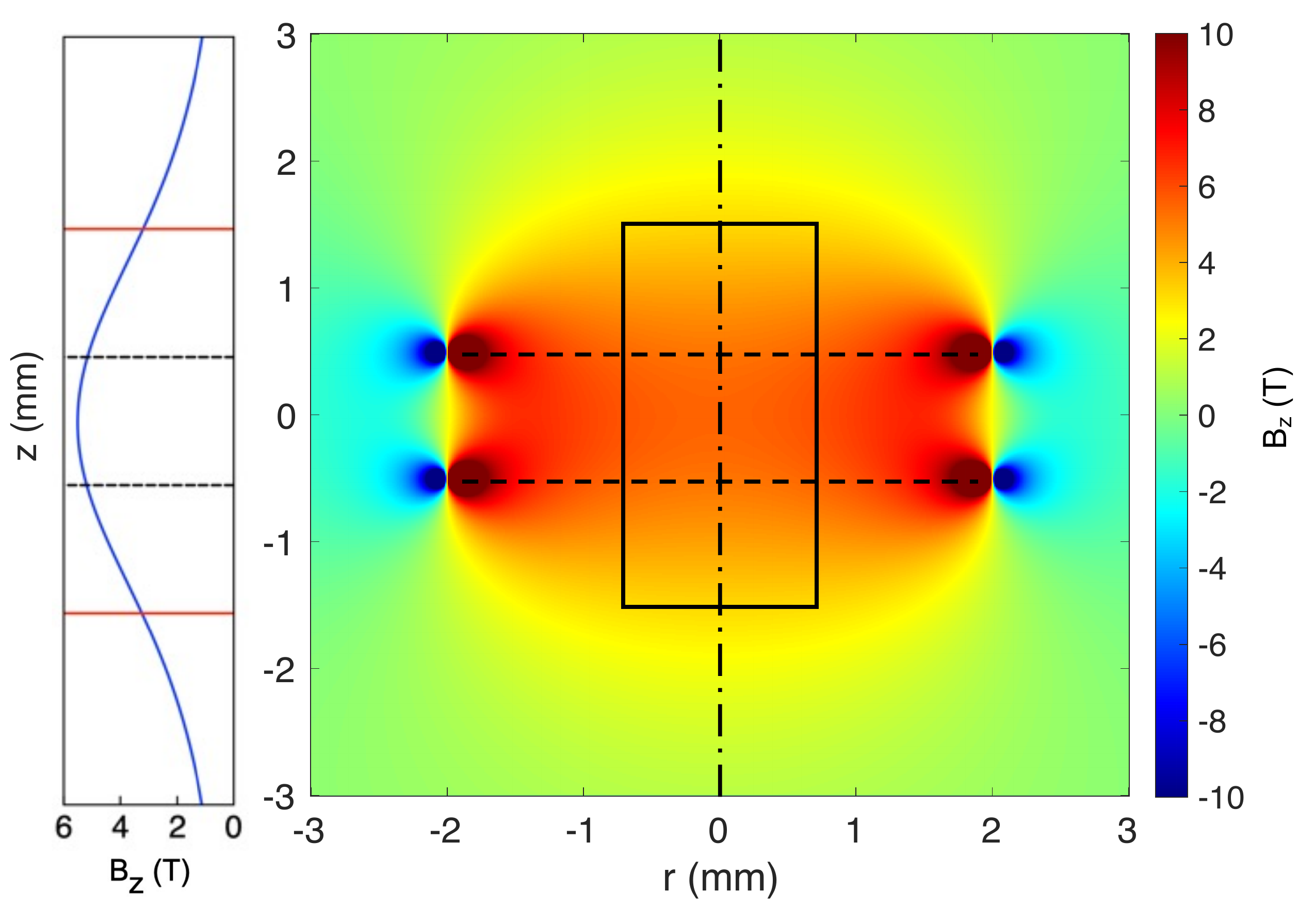}
\caption{\label{fig:B-fieldMap} 2-dimensional distribution of the seed magnetic field across the cylinder. In this figure, the horizontal dashed lines represent the position of the coils, whereas the solid-line rectangle corresponds to a transverse cut of the cylindrical target. The seed magnetic field across the central region of the cylinder is uniform, with a value of $\sim \unit[5.5]{T}$.}
\end{figure}

An example of magnetic field distribution is shown in Figure \ref{fig:B-fieldMap} for the experimental parameters described in Section \ref{sec:ExpConf}, and a pulse length of $\unit[3]{ns}$ (the figure corresponds to the peak B-field obtained at the end of the laser pulse). The solid-lined rectangle corresponds to the cylindrical target, while the horizontal dashed lines represent the position of the coils. A vertical lineout along the cylinder axis (vertical dashed-dotted line) is shown for clarity on the left of the image. The produced B-field has an acceptably uniform value of $\sim \unit[5.5]{T}$ over the length of the central region of the target (between the coils). This field corresponds to a wire current of \unit[9.8]{kA}. Using RADIA \cite{Radia}, the magnetic field can be integrated in a volume around both coil targets to yield a magnetic field energy of 
\begin{equation}
    \int \frac{B^2}{2\mu_0} dV = \unit[1.4]{J}.
\end{equation} 
This corresponds to a fraction of $5\times10^{-5}$ the total laser drive energy of \unit[27]{kJ} used to generate the B-field.

While the rapid increase of the coil B-field could give rise to eddy currents within the target that would oppose the generated seed B-field and preheat the target, this platform is designed to minimize this effect. As mentioned in the previous section, the orientation of the irradiated plates of the laser-driven coil targets is such that the X-ray burst that will be generated will not intersect with the target. Additionally, the laser contrast at \lmj ensures that the intensity of the prepulse is below $\sim\unit[10^7]{W/cm^2}$, which is not sufficient to ionize the target and generate a preplasma. For these reasons, the target should remain an insulator during the rise-time of the B-field, so that no currents can be induced, and the B-field can soak into the cylinder volume.

\section{Implosion hydrodynamics}
\label{sec:Results}

\begin{figure*}
    \centering
    \includegraphics[width=\linewidth]{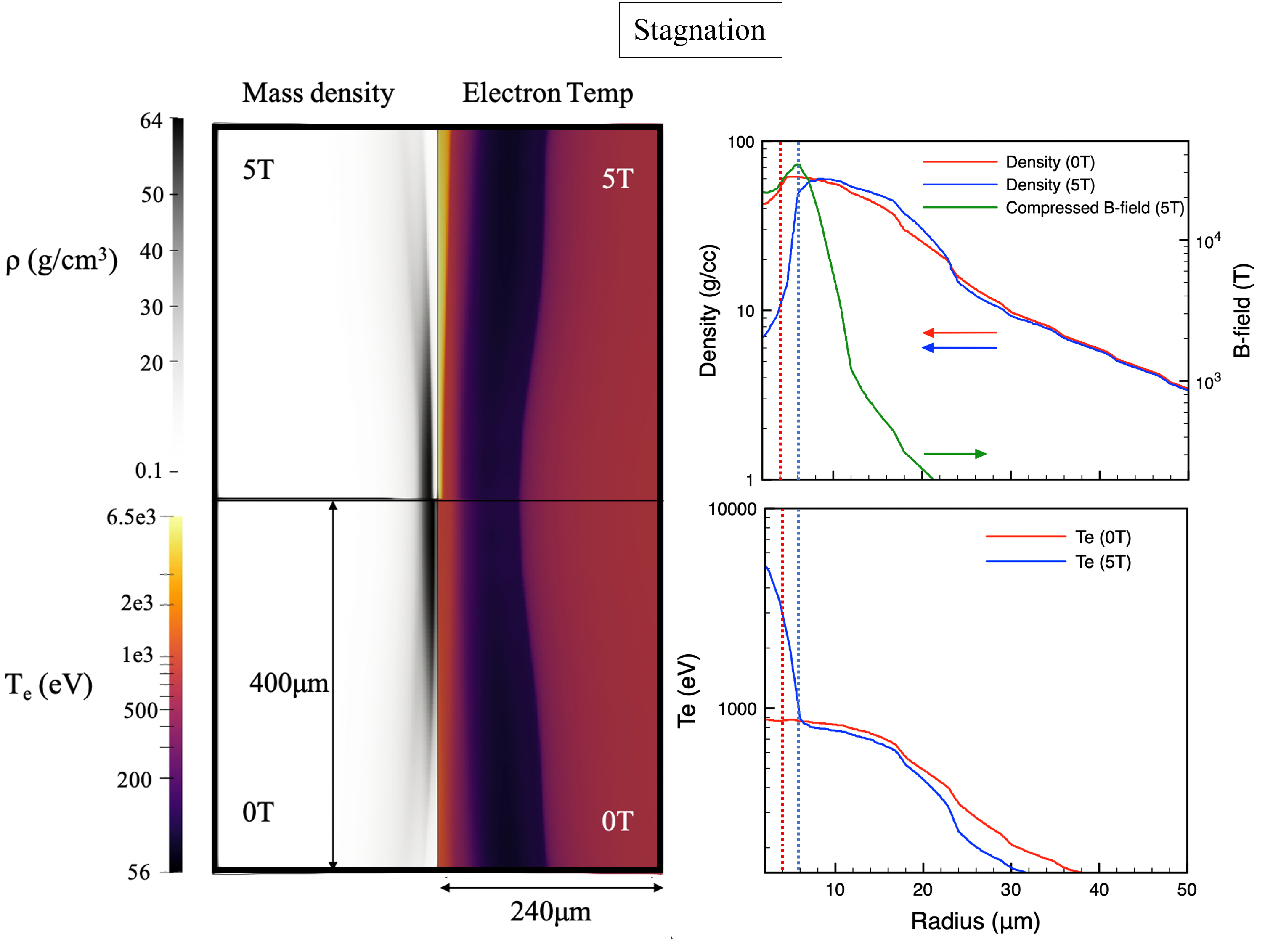}
    \caption{\textbf{Left panel:} 2-dimensional $r-z$ Gorgon simulations showing the density (left side) and temperature (right side) maps at stagnation (\unit[3.5]{ns} after the start of the main drive) for the case with a \unit[5]{T} seed B-field (top) and the unmagnetized case (bottom). \textbf{Right panel:} Radial lineouts through the center of the cylinder, of the density, electron temperature and compressed B-field at stagnation for magnetized (blue) and unmagnetized (red) implosions. The dotted vertical lines indicate the core boundary for both cases. These simulations include radiation transport and Biermann battery effects, along with Nernst and Hall transport.}
    \label{fig:GorgonConditions}
\end{figure*}

To characterize the plasma evolution and measurable outputs from the experiment here proposed, we have performed 2-dimensional extended-\MHD simulations using the code Gorgon \cite{walsh2019, walsh2020, walsh2021}. These simulations include radiation transport, magnetized heat transport, Biermann battery, Nernst effects and updated forms of transport coefficients \cite{sadler2021}, which have been shown to reduce the level of magnetic field twisting for pre-magnetized implosion simulations \cite{walsh2021updated}. We have performed simulations both for a non-magnetized implosion and using a \unit[5]{T} seed B-field, as predicted by the model in Section \ref{sec:coilModel} (see Figure \ref{fig:B-fieldMap}).

In these simulations, a 0.3\% atomic percentage of Ar has been added to fulfill the role of the spectroscopic tracer mentioned in Section \ref{sec:ExpConf}. The addition of this dopant will lower the overall temperature of the implosion, owing to radiative cooling. This effect will depend on the initial magnetic field and the dopant percentage, as discussed in depth in our previous paper \cite{walsh2021}.

The laser-target configuration described in this manuscript is prone to the \CBET parametric instability due to the counter-propagating beams crossing in a large volume and the high intensities involved. To account for this effect, the \MHD simulations presented here include an artificial reduction of the laser energy of 30\%. In order to check this estimation of the \CBET effect, the temperature and density maps obtained from the hydrodynamic simulations were studied with the code IFRIIT \cite{colaitis2019}, including quad-by-quad interaction in the full 3-dimensional configuration at different stages of the implosion. We found high \CBET gains, with reduction in the laser-target coupling up to $\sim 30-40\%$, which validates our first estimate. 

From Figure \ref{fig:laser_imprint}, it can be seen that the laser irradiation pattern presents some modulations (4\% axially and 6\% azimuthally) which could be amplified by \CBET as well. However, the amplification of low-mode modulations by \CBET is a second order effect, with a much lower impact on implosion performance than the reduction of the laser drive itself. Furthermore, recent work on scaling cylindrical implosions to indirect drive facilities has shown how low-mode instabilities can be minimized \cite{sauppe2019, palaniyappan2020}, thus reducing the effect of \CBET-induced growth. For these reasons, we do not expect \CBET amplification of the laser imprint modulations to significantly impact our implosion estimates further than the $30-40\%$ reduction mentioned above. Nevertheless, at the moment, there are three diagnostics at \lmj that can characterize the effects of \CBET, and further \lpi experiments are currently undergoing, so that prior to fielding this platform, these estimates will be updated.

In addition to \CBET, we expect the laser coupling to be reduced further owing to scattered light from \SRS. Nevertheless, the hot electrons generated by this process ($\sim\unit[30-50]{keV}$) are not likely to be detrimental to the implosion performance, since the areal density of the imploding cylinder ($<\rho R> \sim\unit[7-10]{mg/cm^{2}}$) can effectively stop electrons with energies $<\unit[70]{keV}$ from reaching the core. It is possible that more energetic electrons are generated via the \TPD mechanism (up to $\sim \unit[100]{keV}$), or by \SRS in laser filaments, which could eventually preheat the fuel, thus reducing the efficiency of the compression. However, \SRS is the dominant mechanism in the long scale-lengths plasmas expected in our conditions, and the laser intensity in not high enough to lead to significant filamentation and the associated production of high-energy hot electrons through \SRS \cite{rosenberg2018}.

Stagnation occurs at \unit[3.5]{ns} after the start of the main drive in both the unmagnetized case and with a \unit[5]{T} seed B-field. Note that this timescale is different from that shown in Figure \ref{fig:Current_LMJ}, since the \LDC are driven prior to the cylindrical target. Further to this, Figure \ref{fig:Current_LMJ} shows that, after the maximum current is reached, the current stays relatively constant ($\pm 10\%$) for the \unit[3.5]{ns} required to reach stagnation.
The results at stagnation time are shown in Figure \ref{fig:GorgonConditions}. In this figure, the left-side images correspond to 2D maps of the predicted conditions, while the right-side column shows radial lineouts through the center of the cylinder. In the 2-dimensional image, the temperature (right) and density (left) of an unmagnetized (bottom half) and a magnetized (top half) implosion are compared; whereas the images with the lineouts show the radial distribution of the compressed B-field at stagnation and the mass density, at the top; and the electron temperature at the bottom for both the magnetized and the unmagnetized case. The dotted vertical lines indicate the core boundary for both cases.

It can be seen how, in the magnetized case, the core is heated up to $\unit[>5]{keV}$, compared to $\sim$\unit[1]{keV} in the unmagnetized case. On the other hand, when a magnetic field is applied, the density at the center of the core is $\unit[7]{g/cm^3}$, while in the unmagnetized case it is compressed up to $\unit[40]{g/cm^3}$. The former is owed to the fact that the thermal energy losses are reduced, by magnetizing the electrons; while the latter is a consequence of the increased magnetic pressure in the core, as well as the fact that the thermal pressure is increased in a hotter plasma ($P\propto \rho\times T$). Additionally, the initial B-field of $\unit[5]{T}$ is compressed up to $\unit[25]{kT}$ across the core of the target. This corresponds to an amplification of the seed B-field by a factor $\sim 5000$, while the fuel is compressed by a factor of $(R_0/R)^2\sim 10^4$. Note that the B-field seems to be almost \textit{frozen} in the plasma flow (the magnetic flux is conserved throughout the implosion). This is not an approximation made in the simulations, but rather a direct consequence of the fact that the plasma compresses the B-field faster than it can diffuse away. This is characterized by a high value of the \textit{magnetic Reynolds number}, which is defined as
\begin{equation}
R_M=\frac{U R}{\eta}
\end{equation}
where $U$ is the implosion velocity and $R$ is the plasma radius at a given time and $\eta$ is the magnetic diffusivity. From our \MHD simulations we calculate $R_M\sim200$.

To quantify the effect of the magnetic field on the plasma, we use the following metrics \cite{walsh2021}
\begin{equation}
    \beta = \frac{P_{thermal}}{P_{magnetic}},
\end{equation}
\begin{equation}
    \chi_e = \omega_e \cdot \tau_e,
\end{equation}
where $P_{thermal}$ and $P_{magnetic}$ are the thermal and magnetic pressure respectively, $\omega_e$ is the electron cyclotron frequency ($eB/m_e$), and $\tau_e$ is the characteristic timescale of electron-ion collisions ($\propto T_e^{3/2}/n_e$, where $T_e$ and $n_e$ are the electron temperature and density respectively). With these definitions, the $\beta$ parameter indicates the relative significance of the magnetic field for the macroscopic plasma motion (a value of $\beta\lesssim 100$ already means that the magnetic pressure plays a significant role), whereas $\chi_e$ (the so-called \textit{Hall parameter}) gives an indication of the role of the B-field on the electron energy transport (a low $\chi_e$ means low magnetization).

The obtained simulation results in the stagnated core correspond to $\beta\sim 9$ and $\chi_e\sim 40$. This means that the magnetic pressure in the core of the plasma is one ninth of the thermal pressure, and an electron, on average, does $\sim 40$ rotations around the magnetic field line before colliding with an ion. This indicates that a \unit[5]{T} seed B-field is enough to significantly magnetize the implosion, altering electron transport (less energy losses perpendicular to the B-field) and pressure balance, thus modifying the hydrodynamic conditions of the implosion, and of the plasma at stagnation.

The high value of $\chi_e$ indicates that the electrons become magnetized, with Larmor radius smaller than the mean free path. This reduces the energy transport rate due to electron-ion collisions, which can increase electron temperature significantly above the ion temperature. This has two important effects. Firstly, the heat conduction becomes anisotropic since it now occurs preferentially along the electrons' magnetic orbits. This anisotropy not only modifies the temperature profile (as seen in Figure \ref{fig:GorgonConditions}), but also the heat wave propagation, localizing the hot plasma in regions of stronger B-field. Secondly, non-local effects (such as the depletion of the high-energy tail of the electron distribution) are reduced perpendicular to the magnetic field. In addition, large azimuthal currents can be induced, which transport magnetic and thermal energy. The magnetic energy is converted into electron energy through resistive diffusion and Ohmic heating. As an example consequence, the large values of the Hall term indicate that the B-field  itself can be twisted azimuthally owing to these induced currents. Therefore the experimental platform described here opens the path to observing and characterizing these effects, which is fundamental to understand this extreme magnetization regime \cite{walsh2021}.

\section{Extraction of plasma parameters throughout the compression}
\label{sec:diagnostic}

A main goal of the proposed experimental platform is to probe the changes on the hydrodynamic conditions of the imploding core due to the impact of the compressed magnetic field. This can be observed by means of a spectroscopic tracer. For instance, Ar-doping of \ICF implosions is commonly used to extract the  density and temperature conditions of imploding cores \cite{regan2002, florido2011, florido2014, nagayama2014, carpenter2020}. This technique exploits two basic properties of the Ar K-shell spectrum emitted from hot and dense plasmas: (1) the strong dependence on density of the Stark-broadened line shapes, and (2) the dependence (through the atomic population kinetics) of the relative intensity distribution of K-shell lines and associated satellite transitions on electron density and temperature. The Ar K-shell emission is however sensitive to temperatures between $\sim$\unit[600-2500]{eV} and, therefore, for an application to a hotter scenario, a higher-Z spectroscopic tracer, such as Kr, suits better. 

Given the large variations expected for the core conditions in the magnetized case (see Figure \ref{fig:GorgonConditions}), in order to probe the conditions at the different regions of the compressed core we propose the use of a combination of Ar and Kr doping in the deuterium plasma.

\begin{figure*}
    \centering
    \includegraphics[width=\linewidth]{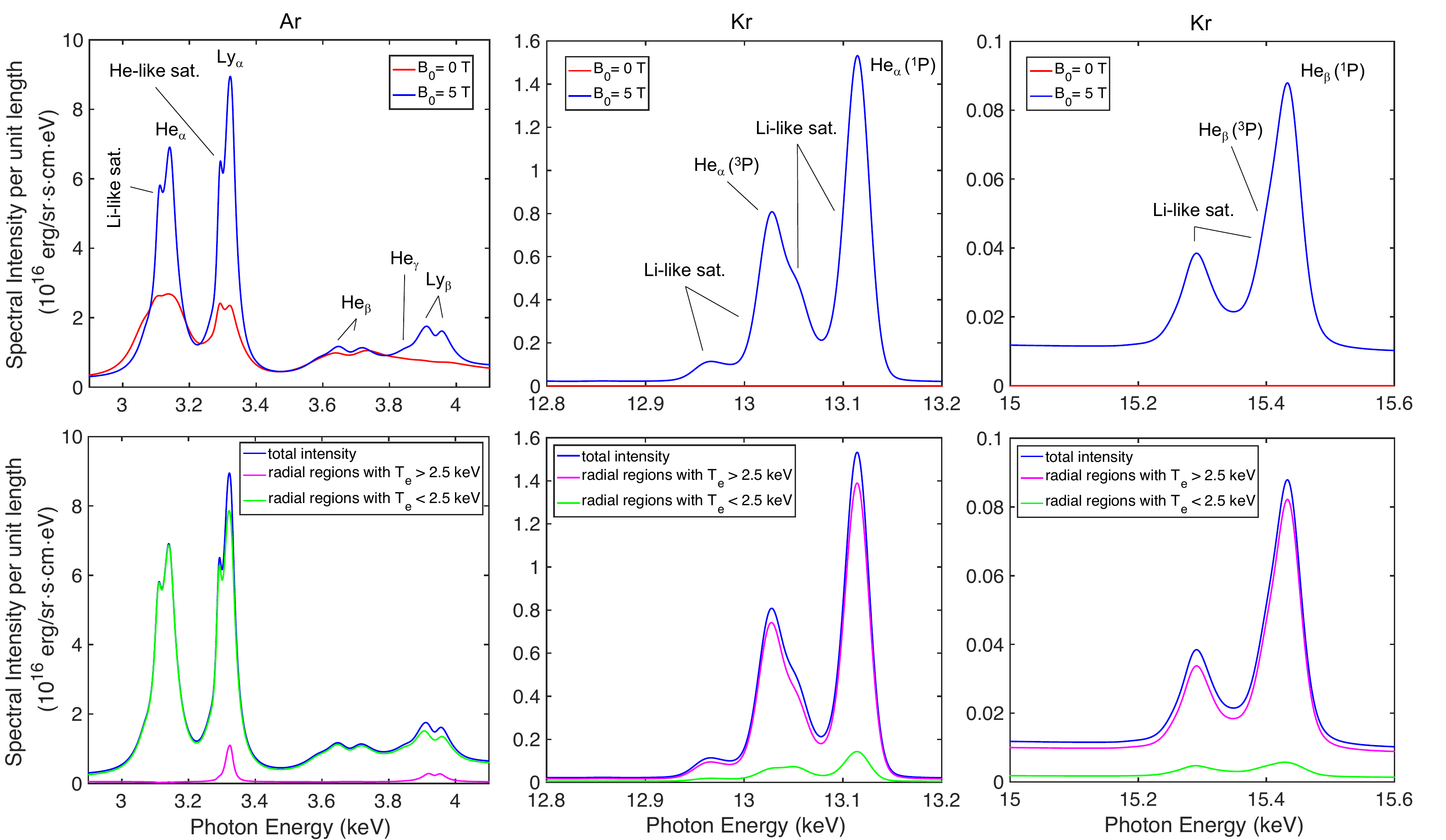}
    \caption{\label{fig:spectra} \textbf{Top row:} Ar (left) and Kr (middle and right) K-shell emission spectra at stagnation, for both a non-magnetized (red) and a magnetized (blue) case. For the Kr case, the middle figure shows the He$_\alpha$ emission region, while the right figure shows the He$_\beta$ lines. These spectra have been obtained using a cylindrical radiation transport model, using the radial profiles of the plasma conditions predicted by Gorgon (as shown in Figure \ref{fig:GorgonConditions}). \textbf{Bottom row:} Ar (left) and Kr (middle and right) K-shell emission spectra at stagnation for the magnetized case. The blue line corresponds to the total intensity, whereas the green and purple lines correspond to the contributions from regions of the plasma at temperatures below and above \unit[2.5]{keV} respectively. This shows that each element probes different regions of the core, introducing an effective spatial resolution. The spectra include contributions from Stark and Doppler broadening, as well as the instrumental broadening corresponding to the resolution achievable at \lmj ($E/\Delta E\sim 500$).}
\end{figure*}

As illustration, the top row of Figure \ref{fig:spectra} shows Ar and Kr K-shell synthetic spectra for the conditions at stagnation for both a non-magnetized (red) and a magnetized (blue) case. Owing to the range of energies that these spectra cover, the Ar K-shell emission (including $n=2\rightarrow1$, $3\rightarrow1$ and $4\rightarrow1$ line transitions in He-like and H-like Ar) is shown on the left, and the Kr He$_\alpha$ ($n=2\rightarrow1$) and He$_\beta$ ($n=3\rightarrow1$) photon energy regions are shown on the middle and right respectively. Consistently with the Gorgon MHD simulations, an Ar atomic concentration of 0.3\% in the deuterium plasma was used. Additionally, based on the study conducted in previous works \cite{chen2017, walsh2021} and in order to keep a low impact of Kr radiative losses on the hydrodynamics, a 0.01\% of Kr was considered. We note in passing that groups of $n=4\rightarrow2$ line transitions in Kr ions --from Be-like to He-like-- also might arise in the Ar K-shell photon energy range shown in Figure \ref{fig:spectra}. However, mainly due to the difference on the concentrations, the referred Kr emission becomes negligible compared to the Ar one.  

These spectra were obtained by solving the radiation transport problem in cylindrical geometry for the radial profiles of temperature and density predicted by Gorgon, as indicated in Figure \ref{fig:GorgonConditions}. Line of sight was assumed to be perpendicular to the cylinder axis. The required frequency-resolved emissivities and opacities (including bound-bound, bound-free and free-free contributions) and atomic level population distributions were calculated using the collisional-radiative model ABAKO \cite{florido2009}. In particular, for this application we used an updated version for multicomponent plasmas. Thus, for given temperature and density values of the plasma mixture, the population kinetics of the Ar and Kr tracers are solved self-consistently, with both species sharing a common free electron pool arising from the ionized deuterium plasma and their own converged ionization balance \cite{sherrill2007}. The attenuation of the core emission through the compressed shell was not explicitly taken into account in the radiation transport calculations presented here. Optical depth estimations for the plastic shell at stagnation conditions in the Ar K-shell photon energy range suggest that a correction to account for the attenuation by the shell --as described in Ref.~\cite{florido2014}-- might be needed in order to properly analyze the spectra and extract the core temperature values. The impact of this effect is expected to be minimum for the higher photon energy range corresponding to the Kr K-shell spectrum. In any case, our estimates indicate that it will be feasible to observe the tracers' spectra. Moreover, previous experiments at the \NIF~\cite{chen2017, gao2022} have observed Kr line emission in similar conditions using \unit[64]{\micro m}-thick plastic shells. 

Importantly for this spectroscopic application, in an attempt to obtain a faithful representation of the emergent spectrum, reliable and detailed Stark-broadened line profiles of the most prominent transitions are needed when performing the radiation transport calculations. In this regard, the Stark line shapes for the shown parent transitions in Ar and Kr -i.e. He$_\alpha$, He$_\beta$, and He$_\gamma$, Ly$_\alpha$ and Ly$_\beta$ in Ar; and He$_\alpha$ and He$_\beta$ in Kr- were obtained by the computer simulation code SIMULA \cite{gigosos2014_A&A}. In this code, the plasma is described as a collection of independent particles trapped in a spherical box and the statistics of relative velocities of the emitters and the perturbing ions are obtained using the so-called \textit{$\mu$-ion model} \cite{seidel1982}. Calculations are then done using the \textit{no-quenching} approximation, i.e. field mixing between the upper (initial) and lower (final) states was neglected due to the large energy separation between them. Furthermore, Stark line profiles of the associated satellite transitions with spectator electron in $n=2$ and $n=3$ were also calculated. For satellite transitions the calculation by the computer simulation technique becomes prohibitive due to the high number of energy states that must be taken into account, and, therefore, the required line shape database was obtained using a recently developed model that follows the framework of the Stark-broadening standard theory \cite{griem1974} and employs an optimized version of the formalism and numerical methods described in Reference \cite{gigosos2014}. We checked that results from this new line shape code agree within $>99\%$ with those obtained using computer simulations with static ions for some selected and affordable cases of interest. It is worth noting that the Zeeman splitting of the lines cannot be used to directly determine the compressed B-field through spectroscopic observation, since in these conditions the Stark broadening is significantly larger (as the electron density reaches values $\sim\unit[10^{25}]{cm^{-3}}$) and blurs the Zeeman pattern \cite{santos2018}. Besides the Stark-broadening mechanism, the spectra shown in Figure \ref{fig:spectra} also include Doppler and instrumental broadening. For the latter, we applied a Gaussian convolution consistent with the spectral resolution achievable at \lmj ($E/\Delta E \sim 500$).  

From Figure \ref{fig:spectra} it can be seen how the krypton acts like a temperature gauge. From this temperature increase, the properties of the B-field may be inferred, since no Kr K-shell emission is observed in the non-magnetized case. Note that, although the Ar emission is still present in the magnetized case, the relative line intensities and shapes are different from the non-magnetized case (e.g., the Ar Ly$_\beta$ line emission is only noticeable in the magnetized case). This is owed to the fact that, while in the non-magnetized case the Ar emission is coming from a roughly uniform plasma at $\sim \unit[1]{keV}$, in the magnetized case, the Ar emission is probing a non-uniform plasma, with temperatures up to $\sim \unit[2.5]{keV}$ (above this value the Ar line emission becomes weak). These changes in the line intensity distribution can be used to extract the differences in temperature and density when the core is magnetized.

The bottom row of Figure \ref{fig:spectra} focuses on the magnetized case, and shows the \textit{effective spatial resolution} that is obtained when using both Ar and Kr as fuel dopants. Most of the emission from Ar comes from regions of the plasma with electron temperatures below $\unit[2.5]{keV}$ (green lines), whereas in the Kr spectra, the main contributions are those from regions with temperatures above this value (purple lines). Therefore, by looking at the corresponding spatial profiles of core conditions at stagnation, it is clear that while the Ar dopant provides information about the core periphery, the Kr emission allows us to gain insight into the conditions at the core center when the fuel is magnetized.

Owing to the capabilities of the currently available spectrometers at \lmj, the emission spectra can be obtained at different times throughout the implosion, providing a temporal map of the temperature and density of the plasma, as described above. If the magnetic field is assumed to be frozen into the plasma motion (which was shown in Section \ref{sec:Results} to be a good approximation) the temperature and density values can be used to estimate a variety of non-dimensional metrics to ultimately determine the relative importance of different magnetization throughout the implosion collapse \cite{walsh2021}.

Additional information can be extracted from the neutron emission from the imploded plasma. According to Gorgon 2D simulations, a 25-fold increase in the neutron yield from the hot-spot is expected for an implosion with a \unit[5]{T} seed B-field, compared to a non-magnetized implosion ($\sim5\times10^{11}$ and $\sim2\times10^{10}$ neutrons respectively). This yield can easily be detected by the neutronic detectors already available at \lmj (where the threshold for neutron detection is $\sim 10^8$)\cite{landoasneutrons}, providing an additional signature of the B-field effects on the implosion.

The implosion velocity and stability can be diagnosed with \xrfc that collect the self-emission from the imploding targets \cite{hansen2018b}. In \lmj, this can be done with both an axial and a radial line of sight, with time resolutions of 110 and \unit[130]{ps}, and spatial resolutions of 35 and \unit[15]{\micro m} respectively \cite{LMJ_GXD}. Additionally, the PETAL beam \cite{blanchot20171} can be used to irradiate a Cu wire, in order to generate an X-ray backlighter and observe the radial profile of the target through X-ray radiography. This is a common technique used in indirect-drive \ICF to diagnose the stability and symmetry of the implosion \cite{rygg2014,dewald2018}.

Finally, diagnosing the seed magnetic field is crucial for understanding the conditions of the experiment. For this purpose we propose the use of proton deflectometry, in a shot with no gas cylinder (only coil targets), using the PETAL beam \cite{blanchot20171} to produce and accelerate protons up to $\unit[51]{MeV}$, via the \TNSA mechanism \cite{raffestin2021}. By placing a reference mesh in the protons' path, and recording the imprint of the beam after it traverses the region between the coils, it is possible to obtain an `image' of the deflections caused by the electric and magnetic fields around the coil targets. Although axial probing of the coil targets (that is, sending the protons along the axis of the coils) has been discussed as an accurate method to characterize the generated B-field \cite{bradford2021,peebles2020}, owing to the large size and inductance of the targets presented here, the B-field signatures that would appear in on-axis radiography are indiscernible with the resolution available at \lmj \cite{bott2021inefficient}. For this reason, we propose probing the targets perpendicular to their axes \cite{santos2018,bradford2020,bradford2021}.

\begin{figure}
\includegraphics[height=0.4\textwidth]{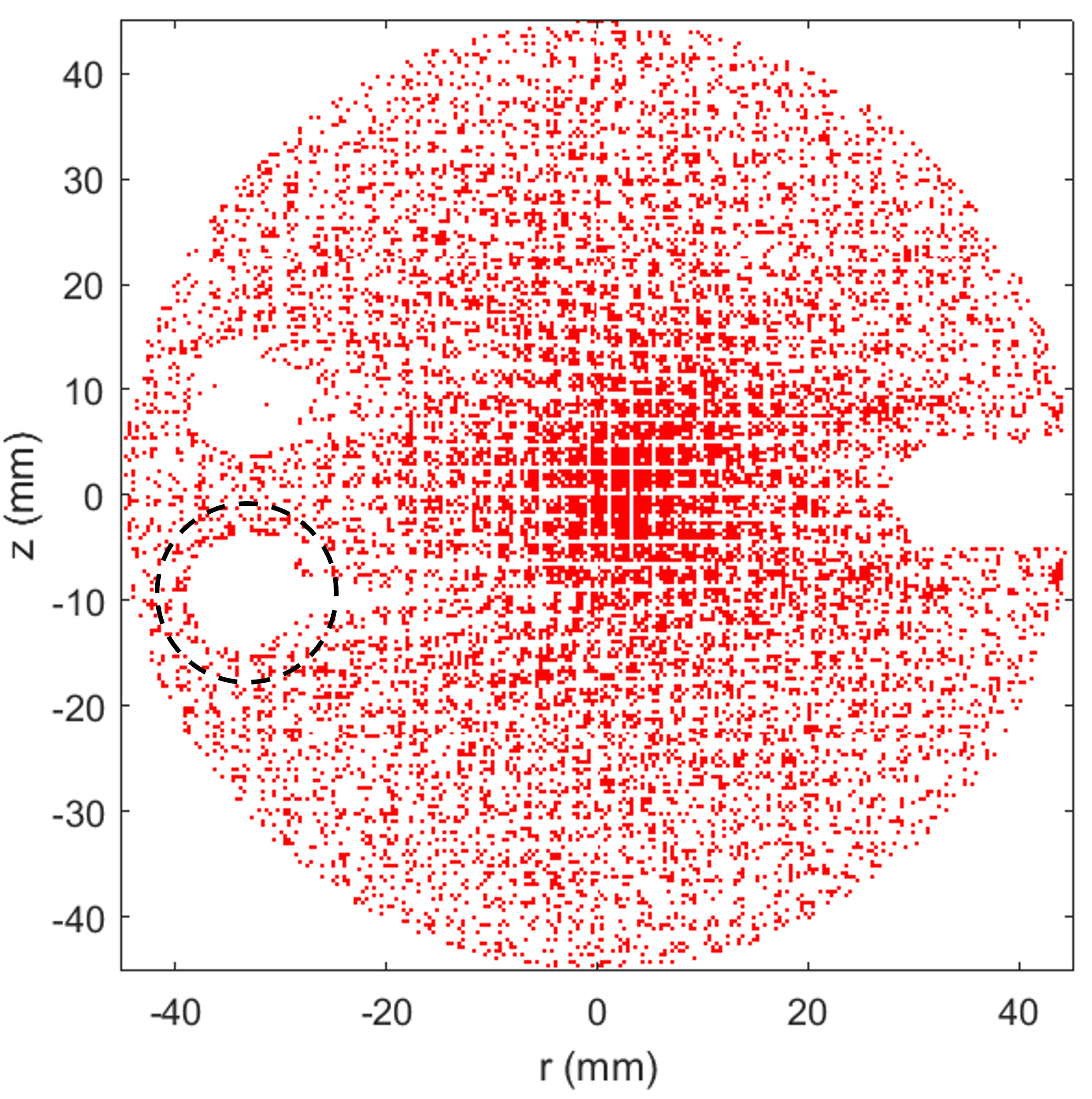}
\caption{\label{fig:ProtonRad} Synthetic proton radiography image obtained with the particle-tracing code PAFIN \cite{pafinThesis} for \unit[14.7]{MeV} protons, using the detector geometry available at \lmj. In particular, the feature that can be observed at the right of the image around $z=0$ corresponds to the radiochromic film (RCF) holder. The magnetic field produced by each coil leaves a characteristic bulb-like feature on the image, and distorts the imprint of the reference mesh (a dashed black circle has been added to point on one of these bulbs). The spatial scale units on the figure correspond to those measured at the detector, so the coils' dimensions are magnified by a factor of $M_c=16$ ensuing proton probing point projection.}
\end{figure}

Figure \ref{fig:ProtonRad} shows a synthetic example image of the expected proton radiography results, obtained with the code PAFIN \cite{pafinThesis} for \unit[14.7]{MeV} protons. In this case, the proton source is \unit[10]{mm} away from \TCC, the reference mesh has a periodicity of \unit[600]{lines/inch} and is placed at \unit[3]{mm} from the source. The detector is placed at \unit[150]{mm} from \TCC, which results in a system magnification of the coil region of $M_c=16$, while the mesh magnification is $M_m\sim 53$. Additionally, the geometry of the \rcf proton detectors at \lmj has been used. This can be seen as the \unit[95]{mm} diameter circular mask, and the white feature at $r\sim\unit[40]{mm}$, $z\sim \unit[0]{mm}$ (which corresponds to the position of the \rcf holder). To produce this figure, a current of \unit[9.8]{kA} looping through the coils is assumed, which corresponds to the peak current obtained with the model introduced in Section \ref{sec:coilModel} and detailed in the Appendix, for a \unit[3]{ns} square pulse with an intensity of $\unit[4\times 10^{15}]{W cm^{-2}}$ (as described in Section \ref{sec:ExpConf}). The electrostatic charge accumulated in the targets can be estimated to be of the order of few nC. Given the size of the laser-driven coil targets presented here, the resulting charge density is sufficiently low so that electrostatic effects cannot be discerned in the proton radiography images. It can be seen how there is a clear imprint of the magnetic field around the coils on the proton image (this is magnified to $z=\pm\unit[8]{mm}$ on the detector, which corresponds to $z=\pm\unit[0.5]{mm}$ on target). In particular, we have marked with a dashed white circle a characteristic bulb-like void feature, which is indicative of a B-field sufficiently strong to deflect all protons around that region. Its dimensions can be used to quantify the B-field generated by the coil discharge current.


\section{Conclusions and future work}
\label{sec:Conclusions}

We have presented an experimental platform design for studying magnetized cylindrical implosions at \lmj. We propose the use of laser-driven coil targets to generate an initial \unit[5]{T} B-field along the axis of the cylinder. Eighty of the \lmj beams will be used to compress the cylindrical target, while eight additional beams will be used to generate the B-field.

The laser-driven coil targets have been modeled following the \textit{diode} model by Tikhonchuk \textit{et al.} \cite{tikhonchuk2017}, suggesting that a fairly uniform seed B-field of $\sim$\unit[5-12]{T} can be achieved over the central \unit[1]{mm} region of the cylinder. This magnetic field can be experimentally diagnosed by means of proton radiography using the PETAL laser.

We have presented a hydrodynamic analysis of the conditions achievable during the implosion and how they are modified by the magnetic field, as it gets compressed with the target. It has been shown that the initial B-field value can be compressed up to $>$\unit[10]{kT}, in a manner which is consistent with the magnetic field being \textit{frozen} in the plasma flow, as a consequence of the high magnetic Reynolds number. Extended-\MHD simulations show that the temperature and density of the plasma at stagnation are heavily affected by the presence of the magnetic field. By doping the fuel, the effects of the magnetic field can then be detected using X-ray spectroscopy.

Besides X-ray spectroscopy, we have proposed a set of diagnostics and their set-up to characterize the evolution of the implosions and the evolution of the magnetic field. This set-up can be directly implemented in the \lmj facility without the need for further development.

Future work includes improving both the performance and understanding of the coil targets, particularly for the poorly explored parameters of the \lmj laser drive (i.e. $\unit[\sim10]{kJ}$ at 3$\omega$). We expect to benchmark the scaling laws used to estimate the hot electron generation from the laser-target interaction at different laser intensities, and particularly for 3$\omega$ light. Parallel experimental efforts using lasers of more modest energies, are currently being made towards simultaneously measuring the plasma density and the self-generated B-fields close to the irradiated plate, in order to characterize the currents that appear in the plasma. Measuring the target stalk leakage current, as well as the plasma impedance between the plates will also help to understand the operation of the targets, and the time limitations before shorting the coil circuit. Furthermore, recent advances in X-ray characterization techniques will allow for experimental probing of the wire surface plasma sheath, resolving the spatial distribution and time evolution of the current.

A comprehensive study of \CBET mitigation techniques will be performed. It is expected that using lower intensity, longer drive pulses with the same energy, and adapting the targets to achieve similar compression ratios might reduce significantly the \CBET impact on the laser-target coupling.

Prior to \lmj experiments, the analysis of recent similar experiments at OMEGA will be used to benchmark the hydrodynamic simulations, and study the relative weights of different transport mechanisms prior to these experiments, in order to produce more accurate predictions.

As mentioned in the introduction one of the points of interest of applying a magnetic field is the confinement of $\alpha$ particles within the core. While not the purpose of this platform directly (and not yet possible at \lmj), implosions using DT fuel might as well be investigated. For the values presented in this paper, the Larmor radius of the $\alpha$ particles would be larger than the compressed core, resulting in not a significant radial confinement. However, $\alpha$ particles would still be confined along the axis of the cylinder, similarly as in \maglif experiments, given that their mean-free-path is smaller than the length of the cylinder. An estimate of this axial confinement is given by the aspect ratio of the compressed core ($R/L$, radius over length). Following the results presented in this paper, we can estimate a loss fraction of $\sim1.3\%$ for the $\alpha$ particles along the axis of the cylinder, which is comparable with \maglif experiments, where this ratio is of the order of 1\% \cite{sinars2020}. Further in the future, \lmj is expected to reach energies above \unit[1]{MJ}. In this case, it will be possible to drive larger targets, yielding a core radius at stagnation larger than the Larmor radius of $\alpha$ particles, for a convergence ratio similar to the presented setup. 

This platform will be fielded experimentally in 2024-2026. The results from these experiments will help benchmarking the different electron transport and B-field advection/diffusion models used in \MHD codes, thus leading to a more accurate understanding of the different mechanisms affecting the hydrodynamic evolution of highly magnetized \HED plasmas.


\section*{Appendix: Modelling and benchmarking the B-field generation in LDCs}

\begin{figure}
\includegraphics[width=\columnwidth]{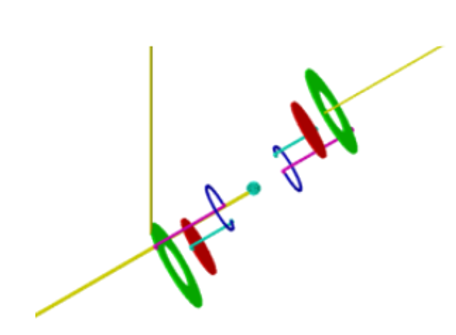}
\caption{\label{fig:VisradOMEGA} Schematic design of the experimental setup used at OMEGA, showing the \LDC and the D$^3$He sphere (light blue in the figure), as seen from the CR39 detector. }
\end{figure}

\begin{figure*}
\centering
\subfloat[\label{fig:OMEGA_Data_3MeV}]{
	\includegraphics[height=0.3\textwidth]{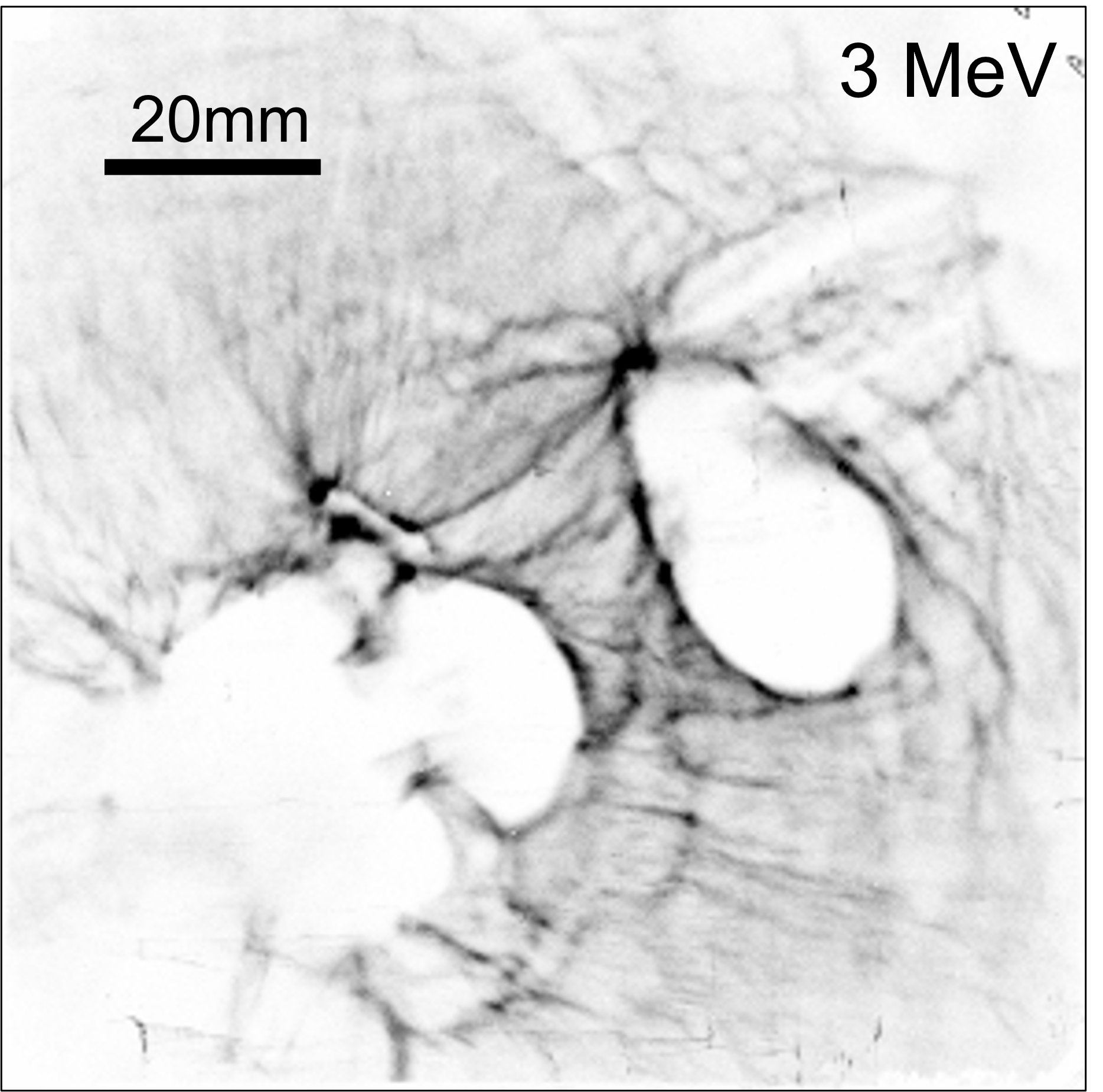}
    }
\subfloat[\label{fig:OMEGA_Model_3MeV}]{
	\includegraphics[height=0.3\textwidth]{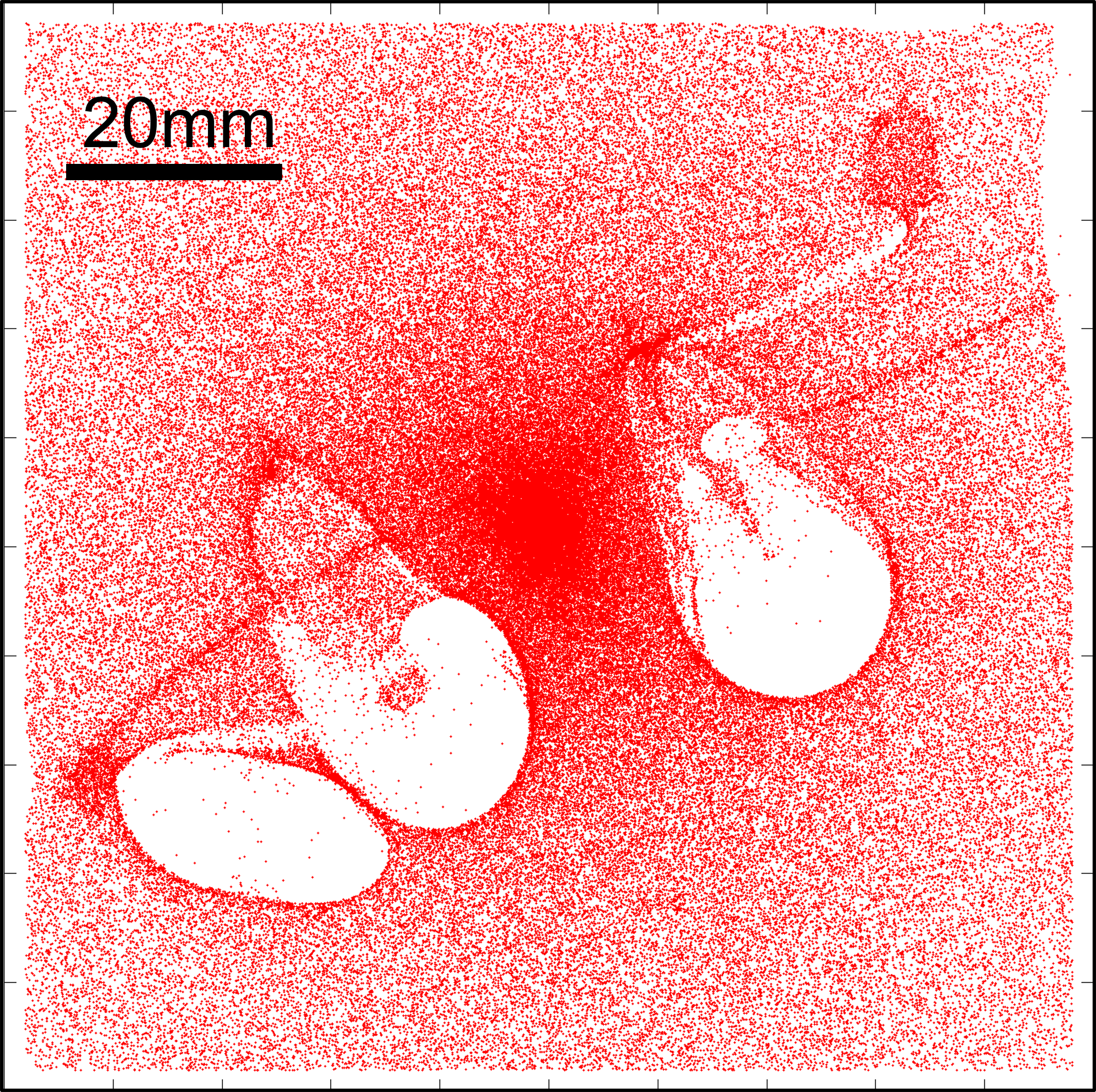}
	}
	\hfill 
\subfloat[\label{fig:OMEGA_Data_15MeV}]{
    \includegraphics[height=0.3\textwidth]{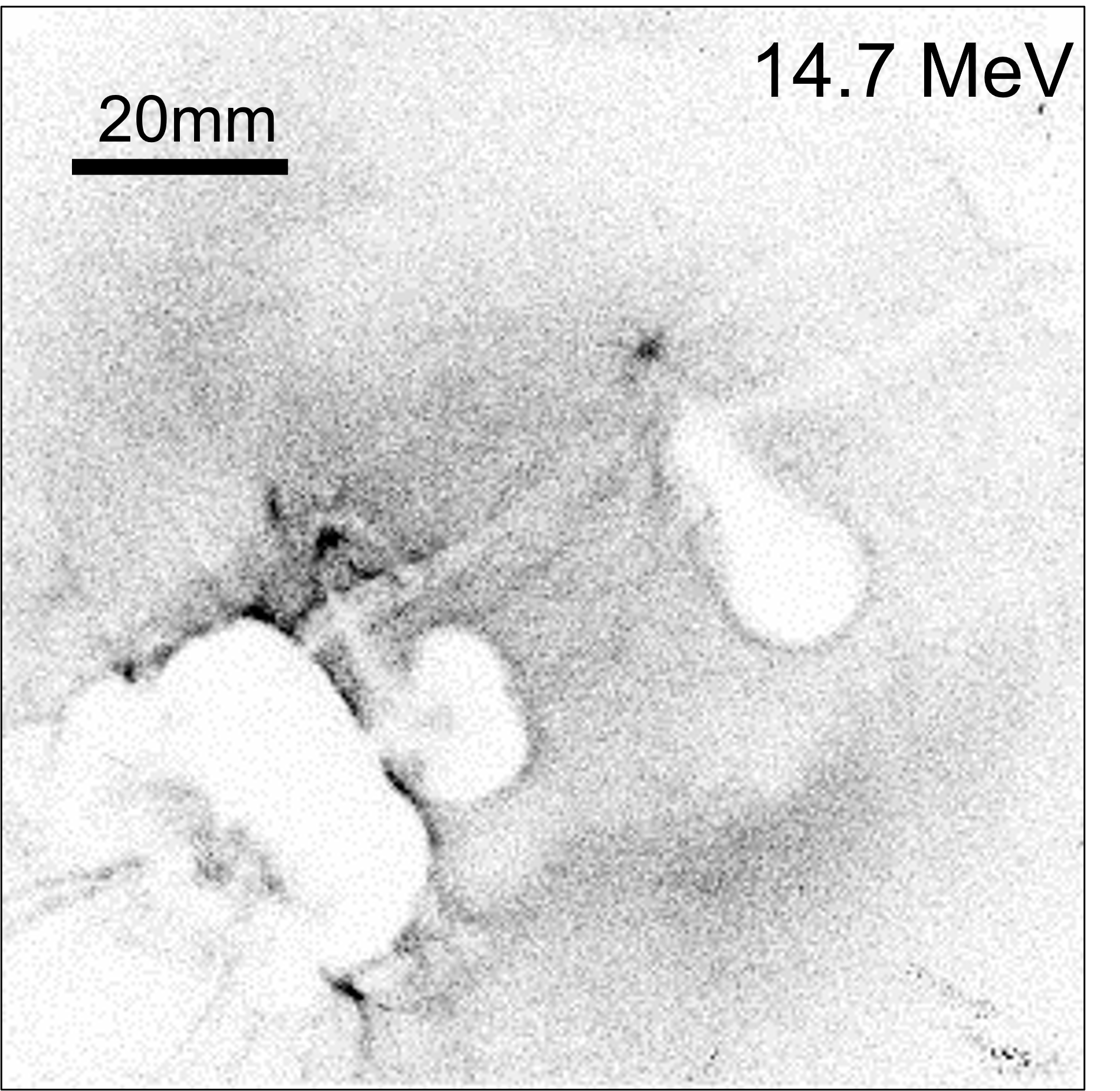}
    }
\subfloat[\label{fig:OMEGA_Model_15MeV}]{
	\includegraphics[height=0.3\textwidth]{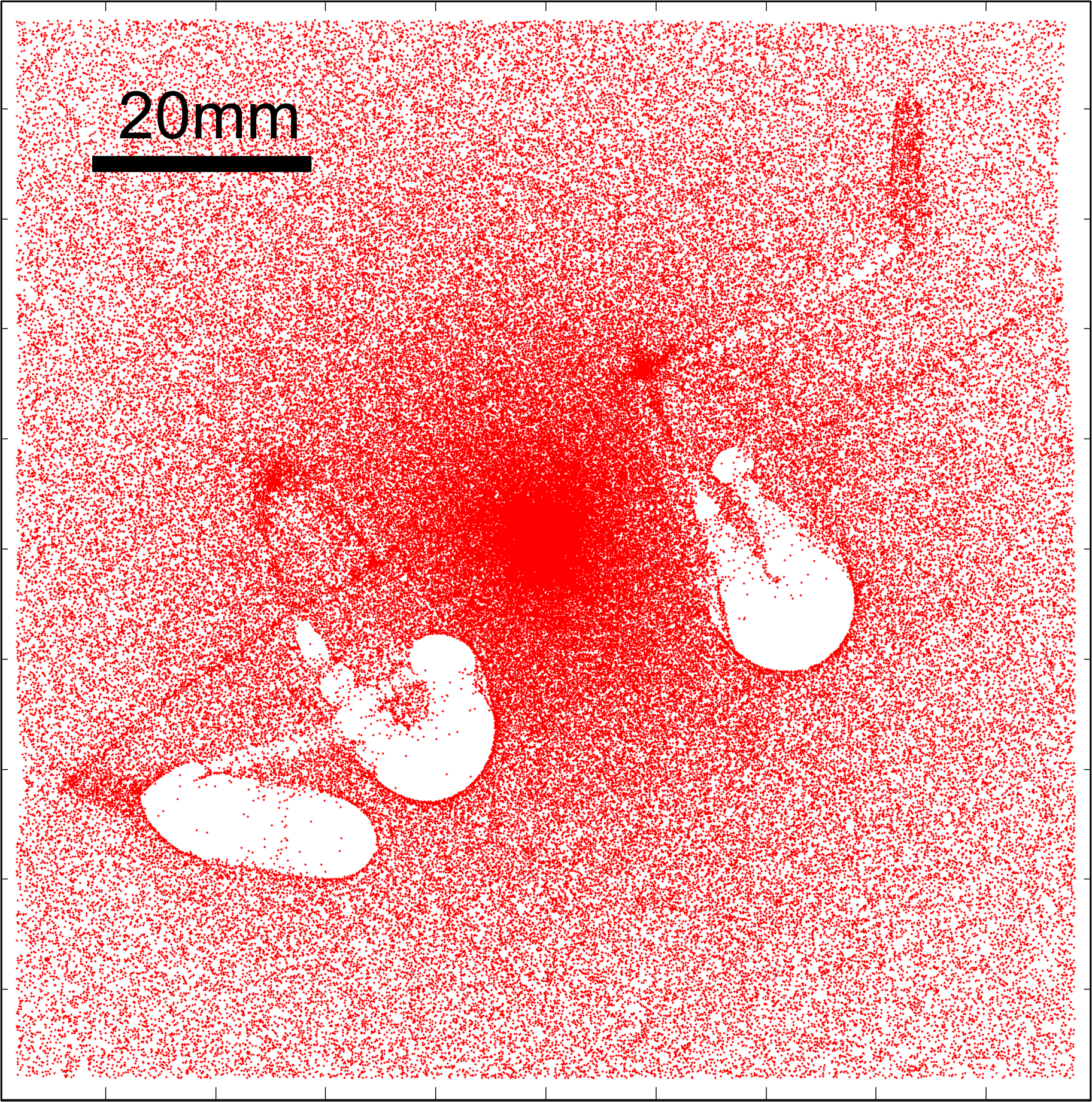}
}
\caption{\label{fig:OMEGA_Radiographs} (\ref{fig:OMEGA_Data_3MeV}) and (\ref{fig:OMEGA_Data_15MeV}) Example of the obtained proton deflectometry images for \unit[3]{MeV} and \unit[14.7]{MeV} protons respectively. (\ref{fig:OMEGA_Model_3MeV} and \ref{fig:OMEGA_Model_15MeV} Corresponding synthetic radiographs produced with the code PAFIN. These synthetic images include currents looping through the targets of \unit[6.5]{kA} and \unit[8]{kA} respectively, as well as additional static toroidal charge distributions around the coils of \unit[+4.5]{nC} and \unit[+8]{nC} respectively.}
\end{figure*}

When a LDC target is irradiated by a focused laser beam, initially, a plasma is generated close to the surface of the irradiated plate. While this plasma has not reached the second plate, the vacuum electron current between the plates is space-charge-limited by the charge that builds up on each plate. When the expanding quasi-neutral plasma reaches the outer plate, which depends on the plasma fast ion velocity, the charge within the plasma is compensated and the potential well close to the interaction plate is strongly reduced. Hot-electron ejection is facilitated and super-Alfvenic ejection currents can be established as the plasma supports a counter-propagating return current. In addition to the space-charge limit, the forward-going current will generate an azimuthal magnetic field that will pinch the plasma between the plates. This magnetization limit depends, among other variables, on the temperature and number of hot electrons.

These two limitations determine the I-V characteristic of the RL-circuit \cite{tikhonchuk2017, williams2020} and, unless the characteristic time $\tau\sim L/R$ (with $R$ and $L$ the circuit resistance and inductance) is reached before the end of the laser pulse, the looping current can increase while the laser keeps irradiating the plate and feeding the system with hot electrons.

The current evolution, $I(t)$, can be computed by solving the following equation,
\begin{equation} 
\label{eq:RL}
V= L \frac{\mathrm{d} I}{\mathrm{d}t}+(Z_{d}+R(t))I,
\end{equation}
where $V$ is the diode potential, $Z_{d}$ is the plasma impedance between the plates (which accounts for both the space-charge and magnetization limits of the plasma diode) and $R(t)$ is the resistance of the external circuit (the coil wire). Most of the system's dynamics are accounted for by the transient resistance in Equation \ref{eq:RL}, which evolves following the I-V characteristic, as a function of the wire temperature. The inductance stays relatively constant for the duration of the laser pulse, given the low expansion velocity of the wire (of the order of \unit[10]{\micro m/ns}, as measured by Santos \textit{et al.} \cite{santos2018}).

Once the current evolution $I(t)$ has been calculated, the 3D spatial distribution of the B-field at any time $t_0$ can be computed for $I(t_0)$ using a magnetostatic code, e.g. RADIA \cite{Radia}, to which the detailed three-dimensional design of the entire LDC has been previously imported. 

There are various physical mechanisms associated with our laser-driven coil platform which cannot be accounted for using such an heuristic plasma-diode model of magnetic field generation. Complications may arise from dense plasma that bridges the target plates and short-circuits the coil, driving the outer plate by prompt X-ray emission from the interaction region, early destruction of the drive plate or deposition of the laser energy in the blow-off plasma before it reaches the solid target.  Maintaining a static current and B-field over the course of a $\unit[\sim3]{ns}$ laser-irradiation time must therefore be seen as a source of risk in these multi-kJ experiments, but previous results on the OMEGA laser, obtained in conditions similar to those at \lmj, give us grounds for optimism.

At OMEGA, we fielded a similar configuration to the one proposed in this manuscript for \lmj, using a pair of \LDC with \unit[750]{\micro m}-radius parallel coils, separated by \unit[2.5]{mm} (this geometry was constrained by the geometry of the OMEGA beams). The inductance of each of the targets was \unit[6.5]{nH}.

Magnetic fields inside the \LDC were characterized using a D$^3$He exploding pusher proton source for deflectometry measurements at two different energies, \unit[3]{MeV} and \unit[14.7]{MeV} \cite{li2008}. Each target was driven simultaneously by independent $3\omega$ laser beams of \unit[2]{kJ} energy and \unit[1.5]{ns} duration focused to $\unit[10^{16}]{W/cm^2}$. This corresponds to an estimated hot electron temperature of $T_h\sim\unit[16]{keV}$ \cite{tikhonchuk2017}. Figure \ref{fig:VisradOMEGA} shows an schematic drawing of the configuration of the \LDC targets at OMEGA, together with the D$^3$He capsule as viewed from the proton deflectometry detector.

Figures \ref{fig:OMEGA_Data_3MeV} and \ref{fig:OMEGA_Data_15MeV} show an example of proton radiography results for \unit[3]{MeV} and \unit[14.7]{MeV} respectively. Accounting for the respective proton energies and the delay between the laser drivers of the \LDC and the pusher, the corresponding probing times are $\unit[1.73\pm0.15]{ns}$ and $\unit[1.5\pm0.15]{ns}$ respectively, with respect to the start of the drive beams. The coil plane was imaged into a CR39 detector with a magnification of $M_c=16$ (the spatial scale given in each image corresponds to the detector plane).

\begin{table*}[]
\caption{\label{table:OMEGA_Results} Results obtained at OMEGA for the two different proton energies, compared with the  predictions from the diode model. The last column includes the minimum and maximum currents predicted within the time uncertainty of the measurements.}
\begin{tabularx}{0.9\textwidth}{>{\centering\arraybackslash}X >{\centering\arraybackslash}X >{\centering\arraybackslash}X >{\centering\arraybackslash}X}
Proton energy  (MeV) & Time (ns) & Measured current (kA) & Predicted current (min. - max. in kA) \\ \hline
 & & &\\
14.7 & $1.50\pm0.15$ & $8\pm 2$ & (10.8 - 11.9)  \\
3 & $1.73\pm0.15$ & $6.5\pm 2$ & (11.4 - 11.8) 
\end{tabularx}
\end{table*}

Teardrop-shaped pinch and void structures are visible around the coils, characteristic of multi-kA currents. We measured the currents and charges present in the \LDC by fitting these images to synthetic radiographs obtained with the particle-tracing code PAFIN \cite{pafinThesis}. An example of these simulations, is shown in Figure \ref{fig:OMEGA_Model_3MeV} for \unit[3]{MeV} protons, and Figure \ref{fig:OMEGA_Model_15MeV} for \unit[14.7]{MeV} protons. These radiographs include a current flowing through the targets (6.5 and \unit[8]{kA} respectively), and a toroidal static charge distributed around the coils themselves (+4.5 and \unit[+8]{nC} respectively). The values of both the current and the charge were adjusted iteratively in order to fit the CR39 data. Note that the features observed in the bottom left corner of the experimental images, are not captured by the synthetic radiographs. These features correspond to the plasma between the plates of the \LDC, which is not included in the particle-tracing PAFIN simulations.

While the value of the current has an effect on the size of the inner proton void in the images, this current alone cannot reproduce the outer caustic feature (the halo around the voids), since this is an effect of the electrostatic charge. These two features (void and halo size) can therefore be used to adjust the current and static charge in the targets separately, with an error of $\unit[\sim2]{kA}$, arising mostly from uncertainties in the target geometry. Table \ref{table:OMEGA_Results} shows the measured currents for both proton energies, compared with the predictions from the described diode model. It can be seen that the model agrees with the measurements (considering the error bars) within a 30\%.

If the current path was significantly deviated from the wire loop, this would still be apparent from the perpendicular radiographs. Additionally, our targets are designed to minimize the ingress of plasma into the coil region, whether from X-ray photoionisation or from the laser focal spot. Moreover, we do not observe extended outer voids/bubbles that could be interpreted as a return current flowing through a surrounding plasma \cite{chien2021}.

These results show that currents of order $\unit[5-10]{kA}$ can be driven in mm-sized coils using $3\omega$ light with ns-duration pulses, and give us ground for optimism.

\section*{Acknowledgements}

G. P.-C. acknowledges funding from the French Agence Nationale de la Recherche (ANR-10-IDEX-03-02, ANR-15-CE30-0011), the Conseil Règional Aquitaine (INTALAX) and the Spanish Ministry of Science and Innovation through the \textit{Margarita Salas} funding program. C.V. and V.O.-B. acknowledge the support from the LIGHT S\&T Graduate Program (PIA3 Investment for the Future Program, ANR-17-EURE-0027). F.S.-V. acknowledges funding from The Royal Society (UK) through a University Research Fellowship. 

This work has been carried out within the framework of the EUROfusion Consortium, funded by the European Union via the Euratom Research and Training Programme (Grant Agreements No. 633053 and No. 101052200 — EUROfusion). Views and opinions expressed are however those of the author(s) only and do not necessarily reflect those of the European Union or the European Commission. Neither the European Union nor the European Commission can be held responsible for them. The involved teams have operated within the framework of the Enabling Research Projects: AWP17-ENR-IFE-CEA-02 \textit{Towards a universal Stark-Zeeman code for spectroscopic diagnostics and for integration in transport codes} and AWP21-ENR-IFE.01.CEA \textit{Advancing shock ignition for direct-drive inertial fusion}.

This study has received financial support from the French State in the framework of the Investments for the Future programme IdEx université de Bordeaux / GPR LIGHT.

This material is based upon work supported by the DOE Office of Science Grant No. DE-SC0022250. The work has also been supported by the Research Grant No. CEI2020-FEI02 from the Consejería de Economía, Industria, Comercio y Conocimiento del Gobierno de Canarias; and by Research Grant No. PID2019-108764RB-I00 from the Spanish Ministry of Science and Innovation.

This work was performed under the auspices of the U.S. Department of Energy by Lawrence Livermore National Laboratory under Contract DE-AC52-07NA27344. This document was prepared as an account of work sponsored by an agency of the United States government. Neither the United States government nor Lawrence Livermore National Security, LLC, nor any of their employees makes any warranty, expressed or implied, or assumes any legal liability or responsibility for the accuracy, completeness, or usefulness of any information, apparatus, product, or process disclosed, or represents that its use would not infringe privately owned rights. Reference herein to any specific commercial product, process, or service by trade name, trademark, manufacturer, or otherwise does not necessarily constitute or imply its endorsement, recommendation, or favoring by the United States government or Lawrence Livermore National Security, LLC. The views and opinions of authors expressed herein do not necessarily state or reflect those of the United States government or Lawrence Livermore National Security, LLC, and shall not be used for advertising or product endorsement purposes. 

\bibliographystyle{ieeetr}

\end{document}